\documentclass{IEEEtran}
\usepackage{color,xcolor}
\usepackage{cite}
\usepackage{amsmath,amssymb,amsfonts,amsthm}
\usepackage{algorithmic}
\usepackage{graphicx}
\usepackage{float}
\usepackage{epstopdf}
\usepackage{textcomp}
\usepackage{multirow}
\usepackage{booktabs}
\usepackage{amsmath}
\usepackage{subfigure}
\usepackage{stfloats}
\usepackage{array}
\usepackage{tabularx}
\usepackage{booktabs}
\allowdisplaybreaks[4]
\def\BibTeX{{\rm B\kern-.05em{\sc i\kern-.025em b}\kern-.08em
    T\kern-.1667em\lower.7ex\hbox{E}\kern-.125emX}}
\begin{document}	
	\newcommand\tbbint{{-\mkern -16mu\int}} 
\newcommand\tbint{{\mathchar '26\mkern -14mu\int}} 
\newcommand\dbbint{{-\mkern -19mu\int}} 
\newcommand\dbint{{\mathchar '26\mkern -18mu\int}} 
\newcommand\bint
{ {\mathchoice{\dbint}{\tbint}{\tbint}{\tbint}} 
} 
\newcommand\bbint{ 
	{\mathchoice{\dbbint}{\tbbint}{\tbbint}{\tbbint}} 	
}	
\title{Vector Single-Source Surface Integral Equation for TE Scattering From {Cylindrical Multilayered Objects}}
\author{Zekun Zhu, \IEEEmembership{Graduate Student Member, IEEE}, Xiaochao Zhou, \IEEEmembership{Graduate Student Member, IEEE}, \\ Shunchuan Yang, \IEEEmembership{Member, IEEE}, Zhizhang (David) Chen, \IEEEmembership{Fellow, IEEE}
\thanks{Manuscript received xxx; revised xxx.}
\thanks{This work was supported in part by the National Natural Science Foundation of China through Grant 61801010, Grant 62071125, Grant 61427803, and Grant 61631002, in part by Fundamental Research Funds for the Central Universities. \textit{(Corresponding author: Shunchuan Yang)}}
\thanks{Z. Zhu, and X. Zhou are with the School of Electronic and Information Engineering, Beihang University, Beijing, 100083, China. (e-mail: zekunzhu@buaa.edu.cn, zhouxiaochao@buaa.edu.cn)}
\thanks{S. Yang is with the Research Institute for Frontier Science and the School of Electronic and Information Engineering, Beihang University, Beijing, 100083, China. (e-mail: scyang@buaa.edu.cn)}
\thanks{Z. Chen is with College of Physics and Information Engineering, Fuzhou	University and on leave from the Department of Electrical and Computer Engineering, Dalhousie University, Halifax, NS, Canada B3J 2X4. (e-mail: z.chen@dal.ca)}
}

\maketitle

\begin{abstract}
A single-source surface integral equation (SS-SIE) for transverse electric (TE) scattering from {cylindrical multilayered} objects is proposed in this paper. By incorporating the differential surface admittance operator (DSAO) and recursively applying the surface equivalence theorem from innermost to outermost boundaries, an equivalent model with only electric current density on the outermost boundary can be obtained. In addition, an integration approach is proposed, {where the small argument expansion of the Hankel function is used to evaluate the singular and nearly singular integrals}. Compared with other SIEs, such as the Poggio-Miller-Chang-Harrington-Wu-Tsai (PMCHWT) formulation, {the computational expenditure is reduced for multilayered structures because only a single source is needed on the outermost boundary}. As shown in the numerical results, the proposed method generates only {19\%} of unknowns, uses {26\%} of memory, and requires {29\%} of the CPU time of the PMCHWT formulation. 
\end{abstract}


\section{Introduction}
\label{sec:introduction}
\IEEEPARstart{T}{he} two-dimensional transverse electric (TE) electromagnetic waves find many practical engineering applications, such as microwave imaging \cite{IMAGING}, remote sensing \cite{REMOTESENSING}, and electromagnetic absorbing materials \cite{RADAR}. The method of moments (MoM) is widely used to solve the TE electromagnetic problems. One of the most commonly used surface-integral-equations (SIEs) is the Poggio-Miller-Chang-Harrington-Wu-Tsai (PMCHWT) formulation {[\citen{PMCHWT}, Ch. 4, pp. 160-188]}. In it, dual sources of both electric and magnetic current densities are considered and solved on each interface of different media. Several techniques, like the combined tangential formulation (CTF) \cite{CTF}, the Müller formulation \cite{MULLER},  and the multiple-trace PMCHWT (MT-PMCHWT) \cite{MT-PMCHWT}, are proposed to improve SIE matrix conditioning and computational efficiency. However, when two or more regions are encountered, a large system of linear equations often results.

{Several methods are proposed to decrease the size or dimension of the system}, like single-source formulations without either magnetic currents or electric currents. In [\citen{SVS}]-[\citen{USE2SVS}], a single-source surface-volume-surface (SS-SVS) formulation is proposed to solve the TE scattering problems of penetrable objects. In \cite{GIBC}, a generalized impedance boundary condition (GIBC) is used to model interconnects without magnetic current density. In [\citen{DSAO}], the differential surface admittance operator (DSAO) without a magnetic current is proposed to model two-dimensional rectangular interconnects. It is extended into other practical applications [\citen{DSAOARBSHAPEDCIM}]-[\citen{DSAOEARLY}]. Other forms of single-source methods are also proposed to model penetrable objects [\citen{SINGLESORC1}]-[\citen{SINGLESORC4}].

Besides the above single-source formulations, equivalence principle algorithms (EPAs) \cite{EPA}\cite{EPAHOF} with both electric and magnetic current densities are proposed to solve three-dimensional scattering problems. In \cite{EPAMPA}, the EPA combined with a connection scheme is proposed to model periodic perfectly electric conductors (PECs) embedded in planar multilayer media for TM scattering. Through the recursive application of the boundary conditions to the interfaces between two different planar media, {multi-layer} scattering problems can be efficiently solved. In the EPAs, both electric and magnetic current densities are required on a closed surface as a result of Love's equivalence theorem {[\citen{LOVE}, Ch.12, pp. 653-658]}. However, when objects are fully embedded in multilayers, the methods mentioned above may suffer from efficiency problems due to the rapid increase of unknowns at each interface.

Other efforts are made to improve the accuracy and robustness for nearly singular and singular integrals in the three-dimensional space \cite{subtraction1,subtraction2,subtraction3,3D-Sing1, 3D-Sing2, 3D-Sing3, 3D-Sing4, 3D-Sing5}.  Green's function in the three-dimensional space is ${e^{-jkR}}/({4{\pi}R})$, and in two-dimensional, the Green's function is the Hankel function ${H_{0}^{(2)}}(\cdot)$, which is the infinite summation of polynomials. They all have the singularity. A singularity cancellation approach is proposed to solve the $log(R)$ singularity in \cite{2D-Sing2}, where the entire-domain basis functions are used in the two-dimensional TM analysis. In \cite{Integral1}, the analytical approach for potential integrals of the logarithmic function $log(R)$ is introduced based on the small argument expansion of ${H_{0}^{(2)}}(\cdot)$. In \cite{2D-Sing3}, a mechanical quadrature approach is proposed to handle the singularity for two-dimensional circular PECs. Only the self-interactions are considered to be singular, and the analytical integration is used to calculate the singular integration, as described in  \cite{SVS} \cite{DSAOARBSHAPEDCIM} {[\citen{TraditionalM1}, Ch.3, pp. 43-44]} \cite{TraditionalM2} {[\citen{MOMBOOK}, Ch.6, pp. 125-137}]. However, when the small argument expansion is used, the self-interaction segments may not fully satisfy the small argument condition, or near segments may also needed to be considered. As a result, those approaches may not get accurate enough results. 

\begin{figure}
	\begin{minipage}[h]{0.45\linewidth}
		\centerline{\includegraphics[scale=0.2]{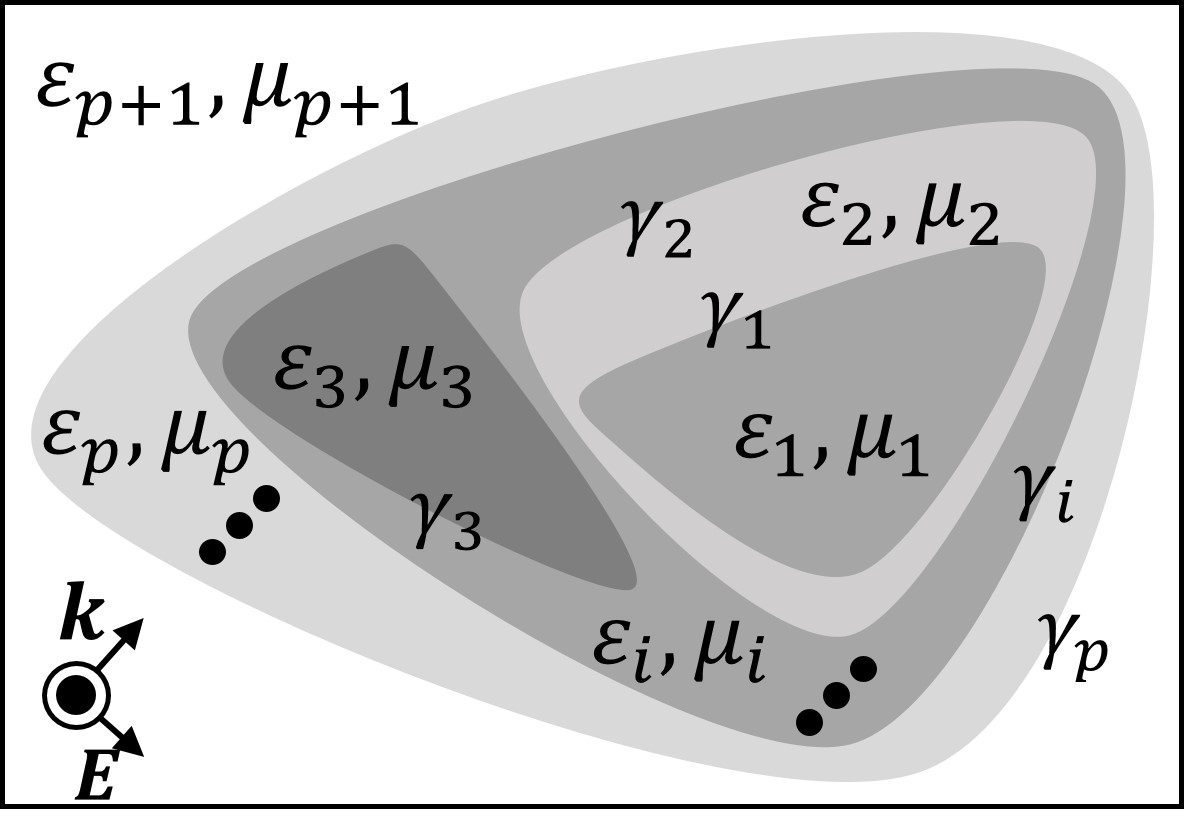}}
		\centerline{(a)}
	\end{minipage}
	\hfill
	\begin{minipage}[h]{0.45\linewidth}
		\centerline{\includegraphics[scale=0.2]{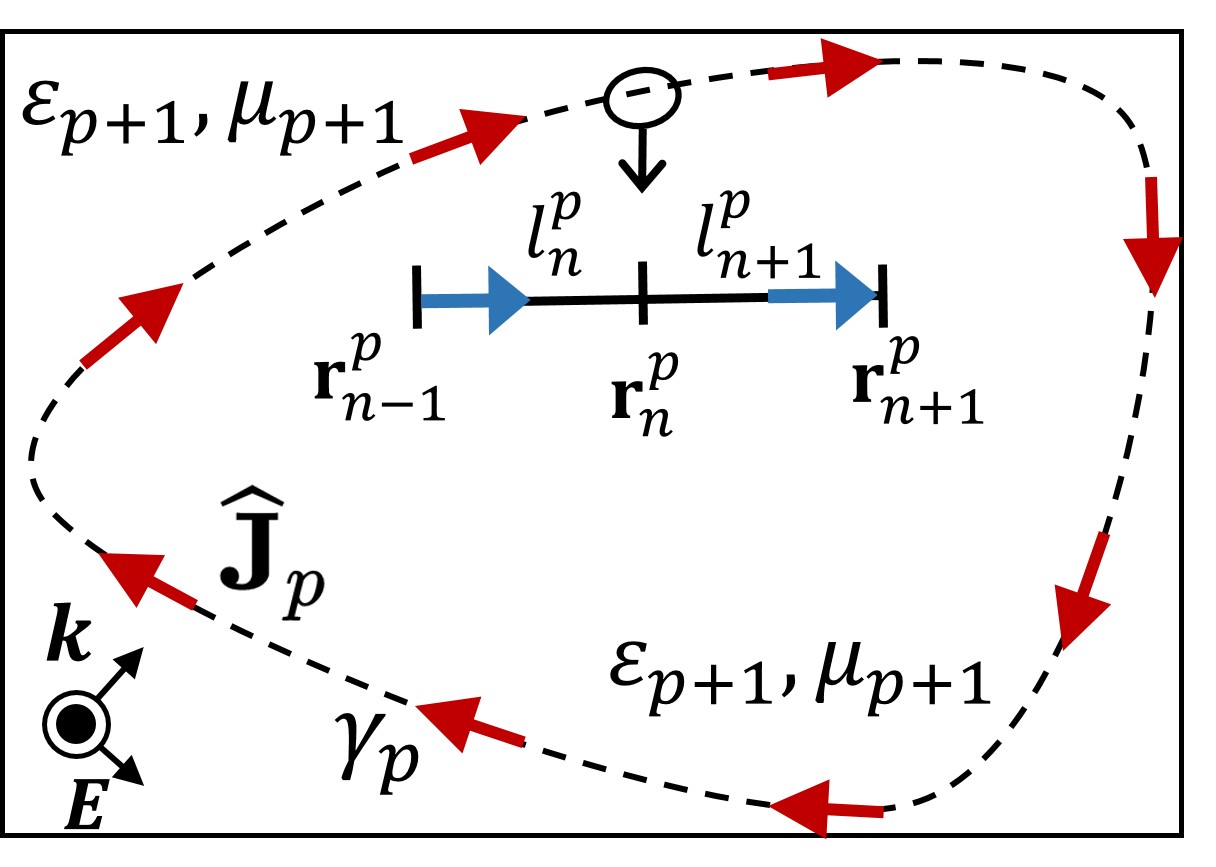}}
		\centerline{(b)}
	\end{minipage}
	\caption{(a) The TE scattering on objects embedded in multilayers. (b) The equivalent model with the electric current density enforced on the outermost boundary $\gamma_i$. }
	\label{MUL_Layer}
\end{figure}

To mitigate the problems described above, a new vector SS-SIE is proposed to solve the TE scattering problems of {cylindrical multilayered objects} (see Fig. \ref{MUL_Layer}(a)). The surface equivalence theorem {[\citen{LOVE}, Ch.12, pp. 653-658]} is recursively applied from the innermost to outermost boundaries, and original objects are replaced by the background medium, as shown in Fig. \ref{MUL_Layer}(b). Furthermore, by incorporating the DSAO \cite{DSAO}, only the equivalent electric current density is needed on the \textit{outermost} boundary, and the computational expenditure is reduced. In addition, an integration approach, {which combines numerical and analytical integration}, is proposed to calculate singular and nearly singular integrals.

It should be noted that the proposed SS-SIE formulation is different from those presented in \cite{RECURSIVE}, in which the Green's function is constructed by recursively uniting fictitious cells with the block inversion of the full-matrix \cite{GIBC}. The proposed formulation directly uses the surface equivalence theorem on each interface and calculates the equivalent current densities. Compared with the conventional methods \cite{GIBC}\cite{RECURSIVE}, the proposed formulation has a physical interpretation rather than purely mathematical manipulations. 

The proposed SS-SIE formulation is also different from that presented in \cite{RECURSIVE2}. In \cite{ RECURSIVE2}, a recursive single-source equation is proposed for scattering by composite objects, in which a pair of exclusive operators are defined to calculate electromagnetic fields. The surface equivalent current densities at interior boundaries are assumed to be related to those on the outermost boundary. In the final system, the current densities on the inner boundaries are to be solved.

{The contributions in this paper are twofold:
	\begin{enumerate}
		\item A new vector SS-SIE formulation is proposed to solve the TE scattering by penetrable or PEC objects embedded in multilayers. {For a complex, multilayered structure, the unknowns on all the boundaries do not have to be solved simultaneously, but are coupled into the outer boundary. It reduces the dimension of the matrix, and thus the computational expenditures.} It is built on our previous work of \cite{MYACES}\cite{MYARXiv}.
		\item An integration approach is proposed to handle various types of singular and nearly singular integrals in the proposed SS-SIE formulation. By combining numerical integration and analytical integration, the matrix elements can be accurately and efficiently calculated. {The number of Gaussian points can be reduced to less than ten to reach the machine level.}
\end{enumerate}}

This paper is organized as follows. In Section II, the proposed SS-SIE formulation is described. In Section III, by combining the surface current density and the electric field integral equation (EFIE) on the outermost boundary, the proposed SS-SIE formulation to solve the exterior scattering problem is presented. In Section IV, {the proposed integration approach} is described. In Section V, the accuracy of {the proposed integration approach} and the performance of the proposed SS-SIE are investigated with several numerical examples. At last, we draw some conclusions in Section VI.
\section{Methodology}
\subsection{The Configurations and Preliminaries}
The TE scattering on {cylindrical multilayered} objects with arbitrary {cross-sections} is considered as shown in Fig. \ref{MUL_Layer}(a). The penetrable object with constant material parameters {$\varepsilon_{i}$ and $\mu_{i}$ is enclosed by its boundary $\gamma_{i}$}. {Assume that there are $p$ layers, and the background medium is with the constant material parameters {$\varepsilon_{p+1}$ and $\mu_{p+1}$}}. Each boundary is discretized with $N_1$, $N_2$, $...$, {$N_p$} line segments. 
Based on the surface equivalence theorem {[\citen{LOVE}, Ch. 12, pp. 653-658]}, the scattering fields can be regarded as generated by surface electric and magnetic current densities on the boundary enclosing the objects. Since the fields outside the objects are of interested, fields inside objects can be \textit{arbitrary} in the equivalent problem. Therefore, the single electric current density ${\widehat{{\mathbf{J}}}_p}$ enforced on the outermost boundary can be obtained through carefully selecting the fields inside objects, as shown in Fig. \ref{MUL_Layer}(b). 

Before we move to detailed formulation derivation, some notations used are introduced. A bold character denotes a vector, and if it has a subscript $i$, the vector is defined on the boundary $\gamma_i$. A hollow character denotes a matrix, and it has the subscript $(i,j)$ and a superscript $(k)$. It means that the source and observation points are on $\gamma_i$ and $\gamma_j$, respectively. $k$ means that the surface equivalence theorem is applied for the $k$th time. A quantity with a wide hat $\ \widehat{}\ $ denotes that it is for the equivalent configuration. {An inner product is defined as}
\begin{equation}
	{\left\langle\mathbf{f},\mathbf{g}\right\rangle = \int \mathbf{f}\cdot\mathbf{g}\ d\mathbf{r}.}
\end{equation}

\subsection{The SS-SIE for A Penetrable Object Embedded in the Background Medium}
A single penetrable object is first considered. We reported the preliminary idea in \cite{APSPAPER}. Detailed formulations are presented in this subsection.

According to the Stratton-Chu formulation {[\citen{Stratton}, Ch. 1, pp. 38-48]}, the electric field $\mathbf{E}$ inside {$\gamma_i$} can be expressed as
\begin{equation}\label{STRACHU}
	{T{\bar{\mathbf{t}}}_{i}({\mathbf{E}}) = {\bar{\mathbf{t}}}_i\left ( {\mathcal{L }}_i [{\mathbf{n}_{i}^{\prime}} \times {\mathbf{H}_{i}}] \right ) + {\bar{\mathbf{t}}}_i\left({\mathcal{K}}_i[{\mathbf{n}_{i}^{\prime}} \times {\mathbf{E}_{i}}] \right),}
\end{equation}
where
{\setlength\abovedisplayskip{10pt}
	\setlength\belowdisplayskip{10pt}
	\begin{align}
		&{{{\bar{\mathbf{t}}}_i\ ({\mathbf{A}_i})} = {{\mathbf{n}_i}} \times {{\mathbf{n}_i}} \times {\mathbf{A}_i}\left({\mathbf{r}}\right),}\label{TANGT} \\[0.5em]
		\label{LL} &{\mathcal{L}_i\left({\mathbf{A}_i} \right) =   - j\omega \mu {(1 + \frac{1}{{{k_i^2}}}{\nabla }{\nabla }\cdot )}\oint_{\gamma_i} G_i({\mathbf{r}},{{\mathbf{r}}{\mathbf{'}}})   {\mathbf{A}_i}\left({\mathbf{r'}}\right)d{{\mathbf{r}}{\mathbf{'}}}},\\
		\label{KK} &{\mathcal{K}_i\left({\mathbf{A}_i}\right) = {\nabla }\times\bbint_{\gamma_i}  {{\mathbf{A}_i}}\left({\mathbf{r'}}\right)   G_i({\mathbf{r}},{{\mathbf{r}}{\mathbf{'}}})d{{\mathbf{r}}{\mathbf{'}}}}.
\end{align}}{$\mathcal{L}_i$ and $\mathcal{K}_i$} are both linear operators, and {$\mathcal{K}_i$} is the Cauchy principal value without including the residual term. {${{{\mathbf{n}_i}}} $} denotes a unit normal vector pointing to the interior region of {$\gamma_i$}, the Green's function is {$G_i({\mathbf{r}},{{\mathbf{r}}{\mathbf{'}}}) =  - \left({j}/{4}\right)H_0^{(2)}(k_i | {{\mathbf{r}} - {{\mathbf{r}}{\mathbf{'}}}} |)$} inside the penetrable object, and {$H_0^{(2)}(\cdot)$} is the $zero$th order Hankel function of the second kind. $k_i$ represents the wave number inside {$\gamma_i$}.
$T = 1/2$ when $\mathbf{r}$ and $\mathbf{r}'$ are located on the same boundary, otherwise, $T = 1$.

The dual vector basis function {${{\mathbf{f }}^i_n}({{\mathbf{r}}'})$} {[\citen{MOMBOOK}, Ch. 2, pp. 92]} and  {${\mathbf{n}}_{i}^{\prime} \times {{\mathbf{f }}_n^i}({{\mathbf{r}}'})$} are used to expand {${{\mathbf{n}}_{i}^{\prime} \times \mathbf{H}_i}$ and ${{\mathbf{n}}_{i}^{\prime} \times \mathbf{E}_i}$}, which are defined as
{\setlength\abovedisplayskip{10pt}
	\setlength\belowdisplayskip{10pt}
	\begin{equation}\label{VBSISFUNC}
		{{{\mathbf{f }}_n^i}({{\mathbf{r}}{\mathbf{'}}}) = \left\{ \begin{array}{l}
			\frac{{{{\mathbf{r}}{\mathbf{'}}} - {\mathbf{r}}_{n-1}^i}}{{{l_n^i}}},{\kern 1pt} {\kern 1pt} {\kern 1pt} {\kern 1pt} {\kern 1pt} {\kern 1pt} {\kern 1pt} {\kern 1pt} {{\mathbf{r}}{\mathbf{'}}} \in [{\mathbf{r}}_{n-1}^i,{\mathbf{r}}_n^i]\\[0.3em]
			\frac{{{\mathbf{r}}_{n+1}^i - {{\mathbf{r}}{\mathbf{'}}}}}{{{l_{n + 1}^i}}},{\kern 1pt} {\kern 1pt} {\kern 1pt} {\kern 1pt} {\kern 1pt} {\kern 1pt} {\kern 1pt} {\kern 1pt} {{\mathbf{r}}{\mathbf{'}}} \in [{\mathbf{r}}_n^i,{\mathbf{r}}_{n+1}^i]
		\end{array} \right.,}
\end{equation}}{where ${{{\mathbf{f}}_n^i}({{\mathbf{r}}'})}$ is the $n{\rm{th}}$ basis function on $\gamma_i$, ${{l_n^i}}$ and ${{l_{n+1}^i}}$ are the length of the $n{\rm{th}}$ and $(n+1){\rm{th}}$ segments, and ${{\mathbf{r}}_{n-1}^i}$, ${\mathbf{r}}_n^i$, ${\mathbf{r}}_{n+1}^i$ are the endpoints of the $n{\rm{th}}$ and $(n+1){\rm{th}}$ segments} as shown in Fig. \ref{MUL_Layer}(b), respectively. {${{\mathbf{f }}_n^i}({{\mathbf{r}}'})$} mimics the RWG basis function in three-dimensional space, and the total charge density for each adjacent segment pair is zero {[\citen{MOMBOOK}, Ch. 8, pp. 259]}.  
Therefore, {${{\mathbf{n}_i^{\prime}}} \times {\mathbf{E}_i}$ and ${{\mathbf{n}_{i}^{\prime}}} \times {\mathbf{H}_i}$} can be expanded as 
\begin{align}
	&{{{\mathbf{n}_{i}^{\prime}}} \times {\mathbf{E}_i} = \sum {{e^i_n} {{\mathbf{n}_{i}^{\prime}}} \times {{\mathbf{f }}_n^{i}}({{\mathbf{r}}}'),} }\\[0.7em]
	&{{{\mathbf{n}_i^{\prime}}} \times {\mathbf{H}_i} = \sum {{h^i_n}{{\mathbf{f }}_n^i}({{\mathbf{r}}}')}}.
\end{align}
It should be noted that {${{\mathbf{n}_i^{\prime}}} \times {\mathbf{E}_i}$} is expanded through the dual vector basis function {${{\mathbf{n}_{i}^{\prime}}} \times {{\mathbf{f }}_n^{i}}({{\mathbf{r}}}') $}, so that the residual term on the left-hand side (LHS) of (\ref{STRACHU}) can be well tested by {${{\mathbf{f }}_m^i}({{\mathbf{r}}})$} \cite{RamK}.

{When we fix the observation points on} {$\gamma_i$}, and {${{\mathbf{f }}_m^i}({{\mathbf{r}}})$} is used to test (\ref{STRACHU}), the following matrix equation can be obtained
{\setlength\abovedisplayskip{10pt}
	\setlength\belowdisplayskip{10pt}
	\begin{equation}\label{PENJ1}
		{\frac{1}{2}{\mathbb{U}}_{(i,i)}{{\mathbf{E}}_i}} = {{\mathbb{L}}^{(1)}_{(i,i)}}{{{\mathbf{H}}_i}} + {{\mathbb{K}}^{(1)}_{(i,i)}}{{{\mathbf{E}}_i}}.
\end{equation}}The entities of matrices {${\mathbb{U}}_{(i,i)}$}, {${\mathbb{L}}^{(1)}_{(i,i)}$ and ${\mathbb{K}}^{(1)}_{(i,i)}$} are computed as follows
	\begin{align}
		&{{[{{\mathbb{U}}_{(i,i)}}]_{mn}}}=-\left\langle\mathbf{f}_{m},{\bar{\mathbf{t}}\left(\mathbf{f}_{n}\right)}\right\rangle,\notag\\ &{[{{\mathbb{L}}^{(k)}_{(i,j)}}]_{mn}} = -\left\langle\mathbf{f}_{m}, {\bar{\mathbf{t}}\left(\mathcal{L}_i\left(\mathbf{f}_{n}\right)\right)}\right\rangle, \notag\\ \label{EFIE_P}&{{[{{\mathbb{K}}^{(k)}_{(i,j)}}]_{mn}}}=-\left\langle\mathbf{f}_{m}, {\bar{\mathbf{t}}\left(\mathcal{K}_i\left[{\mathbf{n'}_i} \times \mathbf{f}_{n}\right]\right)}\right\rangle,
	\end{align}
{where {${\mathbb{U}}_{(i,i)}$}, {${\mathbb{L}}^{(k)}_{(i,j)}$ and ${\mathbb{K}}^{(k)}_{(i,j)}$} are square matrices with the dimension of $N_i \times N_i$, $N_j \times N_i$, and $N_j \times N_i$.} ${\mathbb{U}}_{(i,i)}$ is defined without the superscript since it only depends on the boundary segments.
{${\mathbf{E}}_i$ and ${\mathbf{H}}_i$} are two column vectors including the expansion coefficients defined as follows
{\setlength\abovedisplayskip{10pt}
	\setlength\belowdisplayskip{10pt}
\begin{align}
\label{E1} {{\mathbf{E}}_i} &= {[e_1^i , \quad e_2^i,  \quad  \cdots, \quad {e_n^i}, \quad \cdots, \quad e_{N_1}^i ]^T},\\
\label{H1} {{\mathbf{H}}_i} &= {[h_1^i, \quad h_2^i , \quad  \cdots, \quad {h_n^i}, \quad \cdots, \quad h_{N_1}^i ]^T},
\end{align}}and {$e^i_n, h^i_n$} denote the {$n$th} coefficients of the basis functions defined on {$\gamma_i$}. By moving the second term on the right-hand side (RHS) of (\ref{PENJ1}) to its LHS and inverting the square coefficient matrix, we get
%
\setcounter{equation}{12}
\begin{equation}
	{{{\mathbf{H}}_i}} = \underbrace{{[{{\mathbb{L}}^{(1)}_{(i,i)}}]^{ - 1}}[\frac{1}{2}{{\mathbb{U}}_{(i,i)}} - {{\mathbb{K}}^{(1)}_{(i,i)}}]}_{{{\mathbb{Y}}_i}} {{\mathbf{E}}_i},
\end{equation}
where {${\mathbb{Y}}_i$} is the surface admittance operator (SAO) \cite{DSAO} for the original problem.

After the surface equivalence theorem is applied, the penetrable object is replaced by its surrounding medium, and surface equivalent current density is introduced to ensure fields in the exterior region unchanged. With a similar procedure above, we obtain 
\begin{equation}\label{EqYs}
	{{{\widehat {\mathbf{H}}_i}} =\underbrace { {[{\widehat {\mathbb{L}}^{(1)}_{(i,i)}}]^{ - 1}}[\frac{1}{2} {{\mathbb{U}}_{(i,i)}} - {\widehat {\mathbb{K}}^{(1)}_{(i,i)}}] }_{{\widehat{\mathbb{Y}}}_i} {{{\mathbf{E}}_i}},}
\end{equation}
where {${\widehat {\mathbb{Y}}_i}$} is the SAO for the equivalent problem, {and $k_{i+1}$ should be used in (\ref{EqYs}), since $\varepsilon_{i}$ and $\mu_{i}$ have been replaced by $\varepsilon_{i+1}$ and $\mu_{i+1}$}. {$\widehat {\mathbf{H}}_{i}$,  $\widehat{\mathbb{L}}^{(1)}_{(i,i)}$, and $\widehat{\mathbb{K}}^{(1)}_{(i,i)}$ with wide hat are calculated in the equivalent configuration. It should be noted that there is no hat for ${\mathbf{E}}_{i}$, since we make the tangential electric fields on $\gamma_{i}$ unchanged in the equivalent configuration.}

Therefore, based on the tangential magnetic boundary condition, the surface equivalent electric current density can be expressed as
\begin{equation}{\label{J1}}
	{\widehat{\mathbf{J}}_i }=  {{\widehat {\mathbf{H}}_i}}-{{{\mathbf{H}}_i}} =\underbrace{({ {\widehat {\mathbb{Y}}_i}} - {{\mathbb{Y}}_i)}}_{{{\mathbb{Y}}_{\gamma_i}} }{{{\mathbf{E}}_i}},
\end{equation}
where {$\widehat{\mathbf{J}}_i$} is the expansion coefficients stored in a column vector when the equivalent current density on  {$\gamma_i$} is expanded by {${{\mathbf{f }}_n^i}({\mathbf{r}}')$}, and {${\mathbb{Y}}_{\gamma_i}$ is the DSAO on $\gamma_i$}.


%
\begin{figure}[H]
	\begin{minipage}{0.3\linewidth}
		\centerline{\includegraphics[scale=0.18]{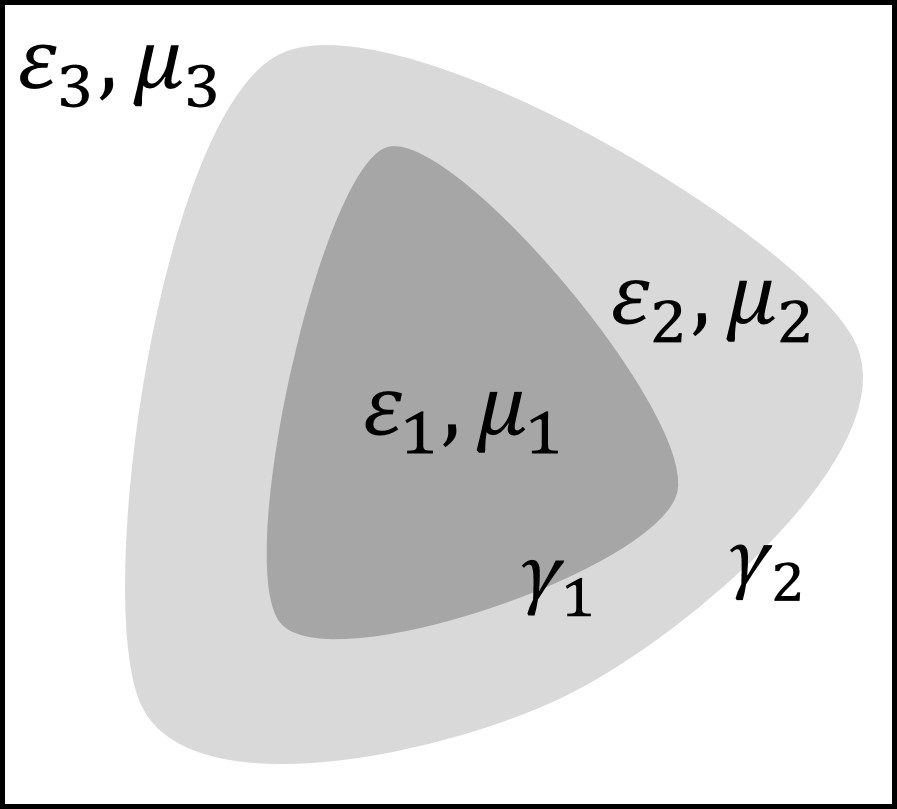}}
		\centerline{(a)}
	\end{minipage}
	\hfill
	\begin{minipage}{0.3\linewidth}
		\centerline{\includegraphics[scale=0.18]{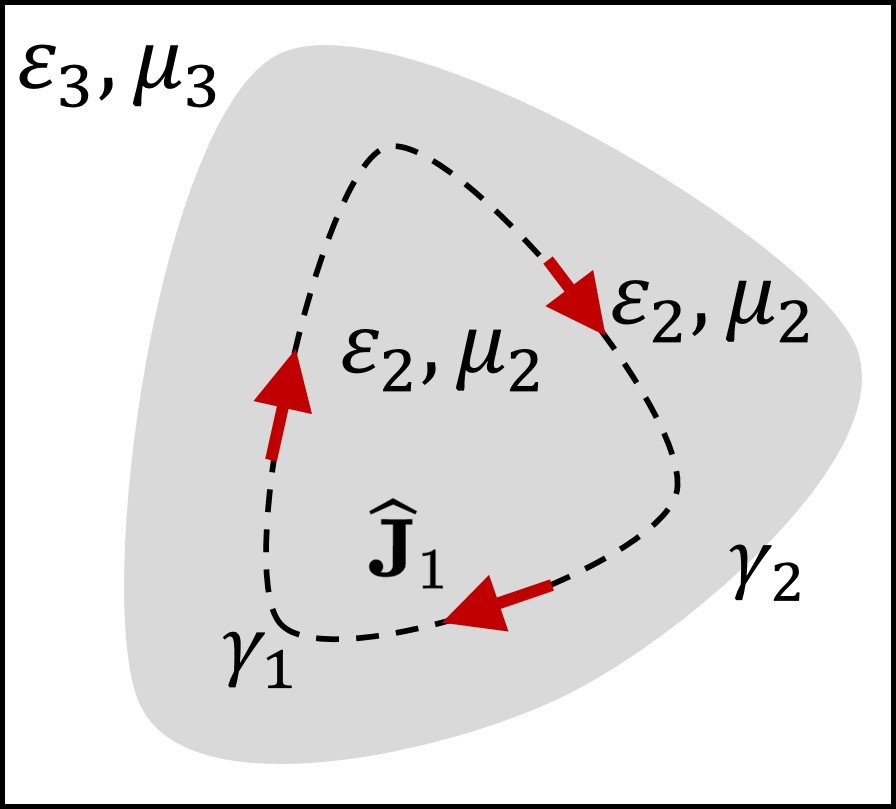}}
		\centerline{(b)}
	\end{minipage}
	\hfill
	\begin{minipage}{0.3\linewidth}
		\centerline{\includegraphics[scale=0.18]{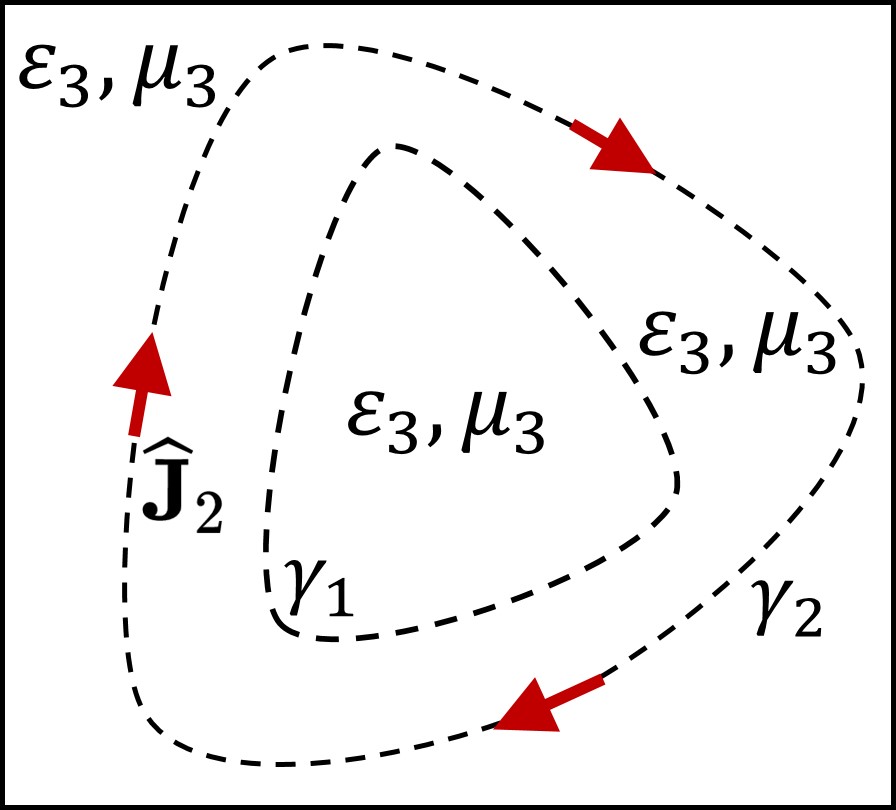}}
		\centerline{(c)}
	\end{minipage}
	\caption{{(a) Two-layered penetrable object is embedded in the background region. (b) The equivalent current density $\widehat{\mathbf{J}}_1$ is introduced on $\gamma_1$ after $\varepsilon_{1}$, $\mu_{1}$ is replaced by $\varepsilon_{2}$, $\mu_{2}$. (c) The equivalent current density $\widehat{\mathbf{J}}_1$ is transferred to $\gamma_2$, and the medium inside $\gamma_2$ is replaced by $\varepsilon_{3}$, $\mu_{3}$.}}
	\label{Two_Layer_O}
\end{figure}

\subsection{The SS-SIE for A Two-layered Object Embedded in the Background Medium}
{Let's consider a two-layered object surrounding by $\gamma_3$, as shown in Fig. \ref{Two_Layer_O}(a)}. Our goal is to derive a surface equivalent electric current density {$\widehat{{\mathbf{J}}}_2$} enforced on the outermost boundary $\gamma_2$ through recursively applying the surface equivalence theorem from $\gamma_1$ to $\gamma_2$. {The proposed SS-SIE formulation is also applicable to the scenario when the PEC objects are embedded inside $\gamma_i$. The derivations are elaborated in the following paragraphs.}

\subsubsection{A Dielectric Object in the Inner Region}
By applying the formulation in Section III-C, the inner penetrable object inside $\gamma_1$ is replaced by its surrounding medium, and a surface equivalent current density {$\widehat{\mathbf{J}}_1$} is obtained on $\gamma_1$, as shown in Fig. \ref{Two_Layer_O}(b). 

{For a homogeneous object with an electric current density, the electric fields on $\gamma_2$ can be expressed through the Stratton-Chu formulation \cite{Stratton} as} 
{\setlength\abovedisplayskip{10pt}
	\setlength\belowdisplayskip{10pt}
\begin{align}
		{T{\bar{\mathbf{t}}}_{i}({\mathbf{E}})} &= {{\bar{\mathbf{t}}}_i\left ( {\mathcal{L }_2} [{\mathbf{n}_2^{\prime}} \times {\mathbf{H}_2(\mathbf{r'})}] \right )}\notag \\
		\label{STRACHU2}&+ {{\bar{\mathbf{t}}}_i\left({\mathcal{K}_2}[{\mathbf{n}_2^{\prime}} \times {\mathbf{E}_2(\mathbf{r'})}] \right) + {\bar{\mathbf{t}}}_i [{\mathcal{L }_1}}({\widehat{\mathbf{J}}_1}){]}.
\end{align}
}{Then, we fix the observation points on $\gamma_1$ to calculate ${\mathbf{E}}_1$ for (\ref{STRACHU2})}. The following matrix equation is obtained	
 {\setlength\abovedisplayskip{10pt}
	\setlength\belowdisplayskip{10pt}
	\begin{equation}{\label{EQ1}}
		{{\mathbb{U}}_{(1,1)}}{{\mathbf{E}}_{\mathbf{1}}} = {{\mathbb{ L}}^{(2)}_{(2,1)}}{{\mathbf{H}}_2} + {{\mathbb{ K}}^{(2)}_{(2,1)}}{{\mathbf{E}}_2} + {{\mathbb{ L}}^{(2)}_{(1,1)}}{\widehat{\mathbf{J}}_1},
	\end{equation}
}{where $\widehat{\mathbf{J}}_1$}, ${\mathbf{E}}_2$, ${{\mathbf{H}}_2}$ denote the coefficient column vectors of the current density on $\gamma_1$ and tangential electric and magnetic fields on $\gamma_2$, respectively. 

By substituting (\ref{J1}) into (\ref{EQ1}), moving the last term on the RHS of (\ref{EQ1}) to its LSH and then inverting the square coefficient matrix, we obtain
\begin{equation}{\label{E2E}}
	{{{\mathbf{E}}_{1}} = {\underbrace{\left[{{\mathbb{U}}_{(1,1)}} - {{\mathbb{L}}^{(2)}_{(1,1)}}{{\mathbb{Y}}_{\gamma_1} }\right]}_{{{\mathbb{C}}_{{\gamma _1}}}}} ^{-1} \left[ {{\mathbb{ L}}^{(2)}_{(2,1)}}{{\mathbf{H}}_2} + {{\mathbb{ K}}^{(2)}_{(2,1)}}{{\mathbf{E}}_2} \right].}
\end{equation}
{Similarly, when the observation points are fixed on $\gamma_2$ to test (\ref{STRACHU2})}, we have
{\setlength\abovedisplayskip{10pt}
	\setlength\belowdisplayskip{10pt}
	\begin{equation}{\label{EQ2}}
		{\frac{1}{2}{{\mathbb{U}}_{(2,2)}}{{\mathbf{E}}_{2}}={{\mathbb{ L}}^{(2)}_{(2,2)}}{{\mathbf{H}}_2}+{{\mathbb{K}}^{(2)}_{(2,2)}}{{\mathbf{E}}_2}+{{\mathbb{L}}^{(2)}_{(1,2)}\widehat{\mathbf{J}}_1}.}
	\end{equation}
} 

By substituting (\ref{J1}) and (\ref{E2E}) into ({\ref{EQ2}}) and rearranging each term, the relationship between the tangential electric and magnetic field on $\gamma_2$ can be expressed as 
\begin{equation}\label{Y2}
	{{{\mathbf{H}}_2} = {{{\mathbb{Y}}_2}}{{\mathbf{E}}_2},}
\end{equation}
where  
\begin{align}
		{{{\mathbb{Y}}_2}}= &{{{\underbrace{\left[ {{{\mathbb{L}}^{(2)}_{(2,2)}} + {{\mathbb{L}}^{(2)}_{(1,2)}}{{\mathbb{Y}}_{{\gamma _1}}}{{\mathbb{C}}_{{\gamma _1}}}{{\mathbb{L}}^{(2)}_{(2,1)}}} \right]}_{{{\mathbb{V}}_{{\gamma _2}}}}}^{ - 1}}}\cdot \notag\\
	\label{2E2E}&{\left[ {\frac{1}{2}{{\mathbb{U}}_{(2,2)}} - {{\mathbb{K}}^{(2)}_{(2,2)}} - {{\mathbb{L}}^{(2)}_{(1,2)}}{{\mathbb{Y}}_{{\gamma _1}}}{{\mathbb{C}}_{{\gamma _1}}}{{\mathbb{K}}^{(2)}_{(2,1)}}} \right]}.
\end{align}
{${{\mathbb{C}}_{\gamma_1}}$ in (\ref{E2E}) and ${{\mathbb{V}}_{\gamma_2}}$ in (\ref{2E2E})} are both the matrices that require to be inverted. We will study the conditioning in Section V.
\subsubsection{A PEC Object in the Innermost Region}
When the innermost embedded object is PEC, {a physical current ${\mathbf{J}}_1$ rather than the equivalent current $\widehat{\mathbf{J}}_1$ exists and the tangential electric fields vanish on $\gamma_1$.} Therefore,  ${\mathbf{E}}_1$ is equal to zero on the LHS of (\ref{EQ1}), and we get
\begin{equation}{\label{EQ1PEC}}
	{{{\mathbf{0}}} = {{\mathbb{ L}}^{(2)}_{(2,1)}}{{\mathbf{H}}_2} + {{\mathbb{ K}}^{(2)}_{(2,1)}}{{\mathbf{E}}_2} + {{\mathbb{L}}^{(2)}_{(1,1)}}{\mathbf{J}}_1.}
\end{equation}
By inverting the square matrix {${\mathbb{L}}^{(2)}_{(1,1)}$}, we get 
\begin{equation}{\label{E2EPEC}}
{	{{\mathbf{J}}_{1}} =-{\left[{{\mathbb{L}}^{(2)}_{(1,1)}} \right]^{-1}} \left[ {{\mathbb{ L}}^{(2)}_{(2,1)}}{{\mathbf{H}}_2} + {{\mathbb{ K}}^{(2)}_{(2,1)}}{{\mathbf{E}}_2} \right].}
\end{equation}
Then, after substituting (\ref{E2EPEC}) into (\ref{EQ2}) and rearranging each term, we have 
\begin{equation}\label{PECY2}
	{{{\mathbf{H}}_2} = {{{\mathbb{Y}}_{\text{PEC}}}}{{\mathbf{E}}_2},}
\end{equation}
where  
\begin{align}
		{{{\mathbb{Y}}_{\text{PEC}}}}= &{{\left[ {{{\mathbb{L}}^{(2)}_{(2,2)}} - {{\mathbb{L}}^{(2)}_{(1,2)}}[{{\mathbb{L}}^{(2)}_{(1,1)}}]^{-1}{{\mathbb{L}}^{(2)}_{(2,1)}}} \right]^{ - 1}}}\cdot \notag\\
		\label{YPEC}&{\left[ {{\frac{1}{2}}{{\mathbb{U}}_{(2,2)}} - {{\mathbb{K}}^{(2)}_{(2,2)}} - {{\mathbb{L}}^{(2)}_{(1,2)}}[{{\mathbb{L}}^{(2)}_{(1,1)}}]^{-1}{{\mathbb{K}}^{(2)}_{(2,1)}}} \right].}
\end{align}

\subsubsection{The Equivalent Problem}
After the surface equivalence theorem is applied on $\gamma_2$, the object is replaced by the background medium, and there is no current source inside $\gamma_2$, as shown in Fig. \ref{Two_Layer_O}(c). {When $k_3$ is used for (\ref{EFIE_P})}, we obtain
\begin{equation}\label{Y2E}
	{{\widehat {\mathbf{H}}_2} = \underbrace{{[{\widehat {\mathbb{L}}^{(2)}_{(2,2)}}]^{ - 1}}[\frac{1}{2} {{\mathbb{U}}_{(2,2)}} - {\widehat {\mathbb{K}}^{(2)}_{(2,2)}}]}_{\widehat {\mathbb{Y}}_2 } {{\mathbf{E}}_2}.}
\end{equation}
The surface electric current density ${\mathbf{J}}_2$ on $\gamma_2$ can be expressed as
\begin{equation}\label{JEQ}
{	{\mathbf{J}}_2 =  {\widehat {\mathbf{H}}_2} - {\mathbf{H}}_2 = \underbrace{({ {\widehat {\mathbb{Y}}_{2}}} - {{\mathbb{Y}}_{2/\text{PEC}}})}_{{\mathbb{Y}}_{\gamma_2}} {{\mathbf{E}}_2},}
\end{equation}
{where ${\mathbb{Y}}_{\gamma_2}$} is the DSAO on $\gamma_2$. {Note that ${\mathbb{Y}}_{2/\text{PEC}}$ is (\ref{2E2E}) for the penetrable object or (\ref{YPEC}) for the PEC object.}

{ Compared with the macromodeling approach proposed in \cite{DSAOARBSHAPEARRAY} for antenna array modeling, which is applicable for the PEC object embedded in the background medium, this paper's approach is applicable for more general scenarios including both penetrable and PEC objects.} In addition, as shown in the later subsection, the proposed SS-SIE formulation is still suitable for objects embedded in multilayers. 

\subsection{Extension of the Proposed Formulation for Objects Embedded in Multilayers}
To generalize the proposed SS-SIE formulation for {cylindrical multilayered objects}, the surface equivalence theorem is recursively applied from $\gamma_1$ to $\gamma_p$ through the procedure in Section II-D. Eventually, the equivalent current density {$\widehat{\mathbf{J}}_p$} on {$\gamma_p$} can be expressed as
\begin{equation}\label{JEQOUT}
		{{\widehat{\mathbf{J}}}_{{p}}} = {{\mathbb{Y}}_{\gamma_p} {{\mathbf{E}}_p}},
\end{equation}
 {where ${\mathbb Y}_{\gamma_p}$} is defined as the DASO on {$\gamma_p$}, which relates the tangential electric field to the surface equivalent current density on {$\gamma_p$}. Detailed recursive formulations are similar to those in \cite{MYARXiv}.

\section{Scattering Modeling of the Proposed SS-SIE}
Once the equivalent current density {$\widehat{\mathbf{J}}_p$} on {$\gamma_p$} is derived, the scattering problem shown in Fig. \ref{MUL_Layer} can be solved by using the EFIE. The total electric field ${{\mathbf{E}}}$ in the exterior region is the superposition of the incident field ${{\mathbf{E}}^{inc}}$ and the scattered field, which is expressed as
\begin{align}
	{{{\mathbf{E}}} \left( {\mathbf{r}} \right) =-j\omega \mu \int_{\gamma _p}}&{{\left(1 + \frac{1}{{{k_{p+1}^2}}}{\nabla {\mathbf{'}}}{\nabla {\mathbf{'}}}\cdot\right)}	 }\notag \\[0.7em]
{\label{SUPERPE}}	&{\widehat{\mathbf{J}}_p\left( {\mathbf{r'}} \right)}{{G_{p+1}}\left( {\mathbf{r}},{\mathbf{r'}} \right) d{\mathbf{r'}} + {{\mathbf{E}}^{inc}}}\left({\mathbf{r}}\right),
\end{align}
where ${\mathbf{r'}}$ is on the outermost boundary {${\gamma_p}$}, and {the wavenumber $k_{p+1}$ in the background medium should be used.
Through fixing the observation points on {$\gamma_p$}}, we have 
\begin{equation}{\label{EFIE}}
	{{{\mathbb{U}}_{(p,p)}} \mathbf{E}_p={{\mathbb{L}}^{(p)}_{(p,p)}} {{\mathbb Y}_{\gamma_p}} {\mathbf{E}}_p +\mathbf{E}^{in}.}
\end{equation}
The entries of $\mathbf{E}^{in}$ are expressed as 
\begin{equation}
	{[\mathbf{E}^{in}]_{m}} = -\left\langle{\mathbf{f}_{m}^{p}}, {{\bar{\mathbf{t}}}_p}\left( {{\mathbf{E}}^{inc}}\right)\right\rangle.
\end{equation}
Therefore, the scattering problem can be solved, and  {${{\mathbf{E}}_p}$} on {${\gamma_p}$} is expressed as
\begin{equation}
		{\mathbf{E}_p=\left[{{\mathbb{U}}_{(p,p)}} - {{\mathbb{L}}^{(p)}_{(p,p)}} {{\mathbb Y}_{\gamma_p}} \right]  ^{-1}\mathbf{E}^{in}.}
\end{equation}
\newcounter{TempEqCnt1} 
\setcounter{TempEqCnt1}{\value{equation}} 
\setcounter{equation}{34} 
\begin{figure*}[hb] 
	\hrulefill  
	\begin{equation}
		\label{L}	{{{\mathbf{f}}_m}({\mathbf{r}}) \cdot  {{\cal L}\left[ {{{\mathbf{f}}_n}\left( {{\mathbf{r'}}} \right)} \right]} {\rm{ }}d{\mathbf{r'}}
			={a_1} {{{\mathbf{f}}_m}({\mathbf{r}}) \cdot \left[ {\left( {{{\mathbf{p}}} - {\mathbf{r}}_{1}^{\prime}} \right)\left( {{I_1} + {I_6}} \right) + {a_4}{I_2} + {I_7}} \right] - } 
			{a_1} \nabla  \cdot {{\mathbf{f}}_m}({\mathbf{r}}) \cdot \left[ {{a_4}{I_1} + {I_6}} \right]}
	\end{equation}
	\begin{equation}\label{K}
		{{{\mathbf{f}}_m}({\mathbf{r}}) \cdot  {{\cal K}\left[ {{\mathbf{n'}} \times {{\mathbf{f}}_n}\left( {{\mathbf{r'}}} \right)} \right]} {\rm{ }}d{\mathbf{r'}} ={{\mathbf{v}}_1}  \left( {{a_3}{I_8} + {a_2}{I_4}} \right) + {{\mathbf{v}}_2}\left( {{a_3}{I_6} + {a_2}{I_3}} \right) + {\mathbf{v}}_3\left( {{a_3}{I_8} + {a_2}{I_5}} \right) + {\mathbf{v}}_4\left( {{a_3}{I_7} + {a_2}{I_4}} \right)}
	\end{equation}
\end{figure*}
\setcounter{equation}{\value{TempEqCnt1}}

\section{Calculation of Singular and Nearly Singular Integrals In the Proposed SS-SIE Formulation}
\begin{figure}[H]
	\centerline{\includegraphics[width=4cm]{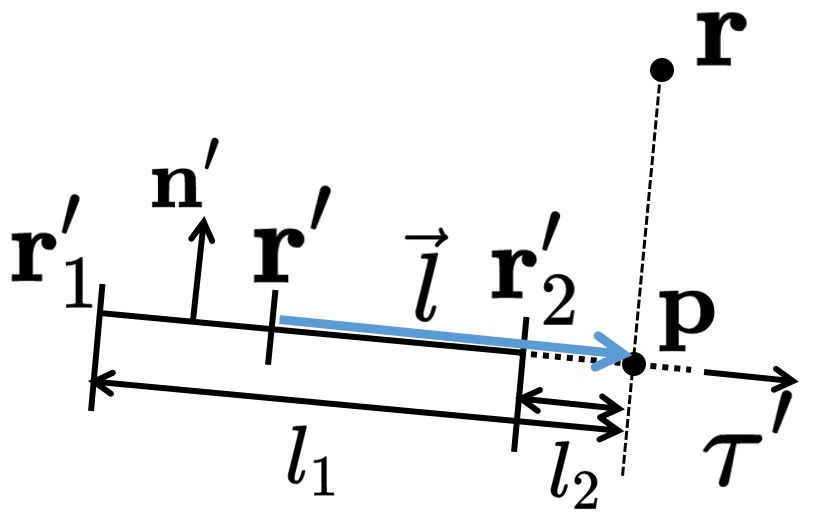}}
	\caption{{A general scenario to demonstrate the geometrical relationship between the observation point and the source segment. $l_1 = (\mathbf{r}_1^{\prime}-\mathbf{r})\cdot\tau^{\prime}$, $l_2 = (\mathbf{r}_2^{\prime}-\mathbf{r})\cdot\tau^{\prime}$, $\vec l = \tau^{\prime}\cdot[(\mathbf{r}^{\prime}-\mathbf{r})\cdot\tau^{\prime}]$.}}
	\label{SC_Configuration}
\end{figure}

{In this Section, a general configuration is presented in Fig. \ref{SC_Configuration} to demonstrate the geometrical relationship between the observation point and the source segment}. The notations in Fig. \ref{SC_Configuration} are the same as those of [\citen{Integral1}]. {$\mathbf{r}_{1}^{\prime}$ and $\mathbf{r}_{2}^{\prime}$ are the endpoints of the source segment $[\mathbf{r}_{1}^{\prime}, \mathbf{r}_{2}^{\prime}]$, $\mathbf{r}$ is the observation point, $\mathbf{r’}$ is the source point located on $ [\mathbf{r}_{1}^{\prime}, \mathbf{r}_{2}^{\prime}]$, $\mathbf{p}$ is the projection point from ${\mathbf{r}}$ to {$(\mathbf{r}_{1}^{\prime}-\mathbf{r}_{2}^{\prime})$}, $\tau^{\prime}$ is the unit vector of {$(\mathbf{r}_{1}^{\prime}-\mathbf{r}_{2}^{\prime})$}, and $\mathbf{n}^{\prime}$ is the unit normal vector.}  

\begin{figure}[H]
	\centerline{\includegraphics[width=5cm]{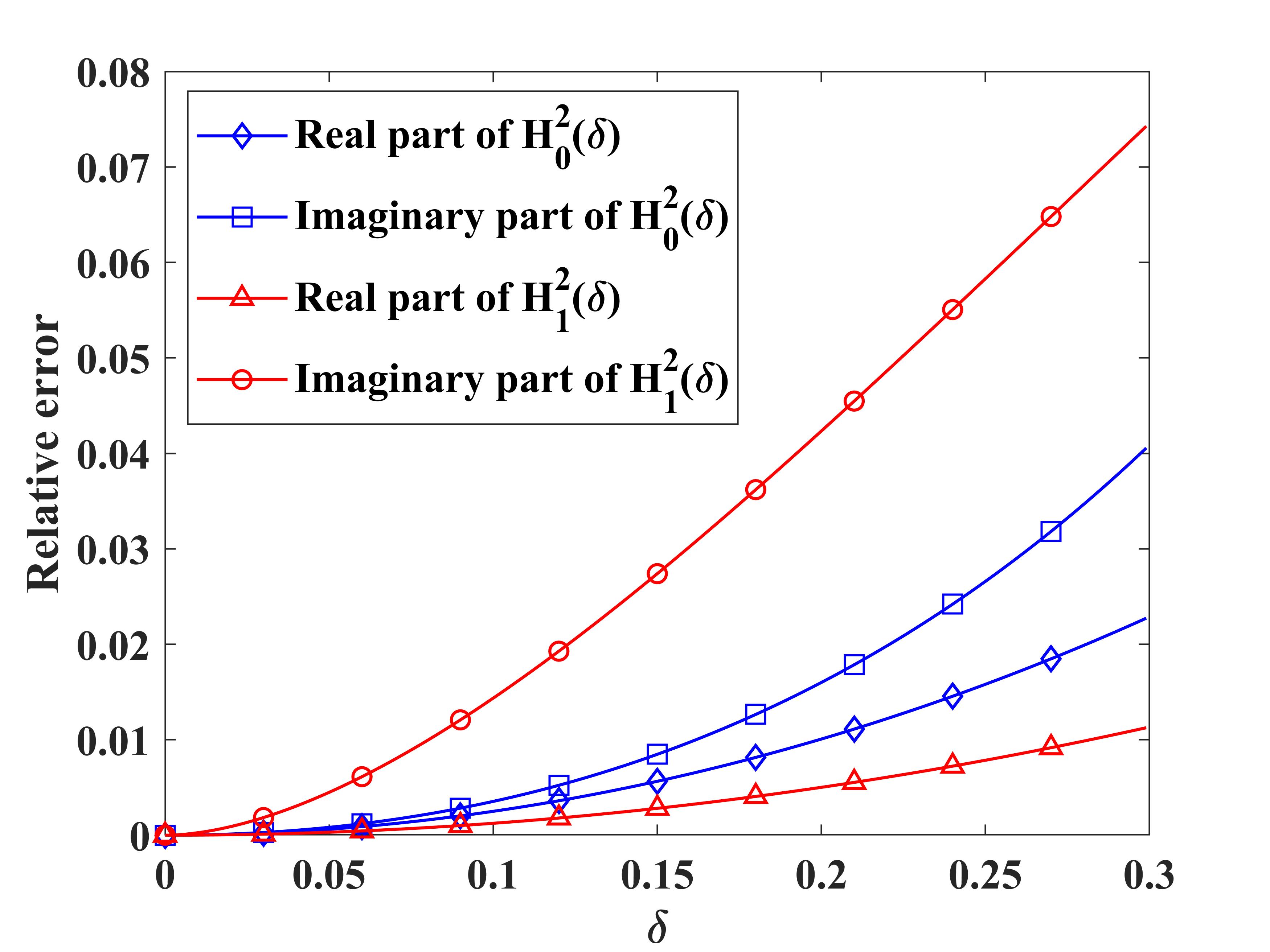}}
	\caption{{The relative error of the small variable expansion compared with Hankle function when different $\delta$ is used. The reference results come from the Matlab command ``besselj(order, $\delta$)" and ``bessely(order, $\delta$)".}}
	\label{curve}
\end{figure}

{For the integration in (\ref{LL}) and (\ref{KK}), the small variable expansion [\citen{Integral1}] is used to analytically calculate the results.} According to [\citen{Integral1}], when $\left|k(\mathbf{r}-\mathbf{r}^{\prime})\right| < \delta$ ($\delta$ is the threshold value to control the accuracy), we use the {first-order approximation of {$H_{0}^{(2)}(\cdot)$} and {$ H_{1}^{(2)}(\cdot)$} to evaluate them}, and the Green’s function and its gradient can be expressed as
\begin{align}
	\label{SmallAforG}
	&G\left(\left|{k(\mathbf{r}-\mathbf{r}^{\prime})}\right|\right) \approx -\frac{j}{4}\left[1-j \frac{2}{\pi} \ln \left(\frac{\gamma \left|{k(\mathbf{r}-\mathbf{r}^{\prime})}\right|}{2}\right)\right], \\
	\label{SmallAforGG}
	&\nabla 'G\left(\left|{k(\mathbf{r}-\mathbf{r}^{\prime})}\right|\right) \approx  - \frac{j}{4}\left({\frac{{{k^2}}}{2} + \frac{{2j}}{{\pi {{\left|{\mathbf{r}-\mathbf{r}^{\prime}}\right|}^2}}}}\right){({\mathbf{r}-\mathbf{r}^{\prime}})},
\end{align}
{where $\gamma$ is the natural exponential of the Euler constant, and 16 effective digits are used in all simulations. Euler constant is approximated as $0.5772156649$. The accuracy of the  approximation is given in Fig. \ref{curve} for various $\delta$. The reference results are obtained through the Matlab command ``besselj(order, $\delta$)" and ``bessely(order, $\delta$)".}

{We assume that all $\mathbf{r'}$ located on $[\mathbf{r}_{1}^{\prime}, \mathbf{r}_{2}^{\prime}]$ satisfy the condition of small variable expansion, $\left|k({\mathbf{r}-\mathbf{r}^{\prime}})\right| < \delta$. When the observation point is fixed at $\mathbf{r}$, and corresponding testing function $\mathbf{f}_m(\mathbf{r})$ is used, the analytical results of $\mathcal{L}$ and $\mathcal{K}$ are expressed as (\ref{L}) and (\ref{K}).}  $a_1$, $a_2$, ${a_3}$, ${a_4}$ are constants defined as
\setcounter{equation}{36}
{
\begin{align}
			&{{a_1} = {\omega \mu }/({{4|{\mathbf{r}^{\prime}_{1}-\mathbf{r}^{\prime}_{2}}|}}),\quad{a_2} = -1/({2{\pi}|{\mathbf{r}^{\prime}_{1}-\mathbf{r}^{\prime}_{2}}|})} ,\notag \\
			&{{a_3} = j({{k^2}}/{8|{\mathbf{r}^{\prime}_{1}-\mathbf{r}^{\prime}_{2}}|}),\ \ {a_4} =  - j({2}/{\pi })\ln (\gamma k/2)},
\end{align}
}and ${{\mathbf{v}}_1}$, ${{\mathbf{v}}_2}$, ${{\mathbf{v}}_3}$, ${{\mathbf{v}}_4}$ are constant vectors defined as
{\begin{align}
			{{\mathbf{v}}_1} = &{{\mathbf{f}}_m \cdot \{[{\mathbf{n'}} \times {({\mathbf{r}}_{1}^{\prime}-{{\mathbf{p}}})}]\times {\tau}^{\prime}\}}, \notag \\
			{{\mathbf{v}}_2} = &{{\mathbf{f}}_m\cdot\{[{\mathbf{n'}} \times {({{{\mathbf{r}}_{1}^{\prime}-\mathbf{p}}})}] \times {({\mathbf{r}}-{{\mathbf{p}}})} \}},\notag \\
			{{\mathbf{v}}_3} = &{{\mathbf{f}}_m \cdot {\mathbf{n'}}},\ \ \ 
			{{\mathbf{v}}_4} = {{\mathbf{f}}_m \cdot {\vec{\tau}}\cdot |{(\mathbf{r}-\mathbf{p})}|}.
\end{align}
}In (\ref{L}) and (\ref{K}), $I_1{\sim}I_8$ are the integrals defined in the Appendix. Here, we use one half of the basis function as a computational example, and the other half can be computed similarly.

{If the Galerkin scheme is used to calculate (\ref{LL}) and (\ref{KK}), there are two integral variables. The outer line integration is defined for $\mathbf{r}$ and the inner line integration is for $\mathbf{r'}$}. {Considering an arbitrary pair of observation and source segments, we proposed an approach to calculate the nearly singular and singular integration in the proposed SS-SIE formulation as described below.} 
\begin{enumerate}
	\item {For the outer line integration, Gaussian quadrature is used to calculate the integrals for $\mathbf{r}$. The integration accuracy will increase when more Gaussian points are used [\citen{GauQua}].}
	\item {For the inner integration, an approach combining the analytical integration approach and Gaussian quadrature is used. When a Gaussian point on the observation segment is fixed, the part of source segment, which satisfies $\left|k{(\mathbf{r}-\mathbf{r}^{\prime})}\right| < \delta$, is analytically calculated by (\ref{L}) and (\ref{K}) (shown at the bottom of this page), and the remaining parts are evaluated by the Gaussian quadrature.}
\end{enumerate}
{The proposed integration approach is applicable not only when the source and observation segments are overlapped, but also for strong nearly singular integrals when the source and observation segments are near to each other.}

\section{Numerical Results and Discussion}


A numerical example is first carried out to verify the convergence of the {proposed integration approach} in Section IV. Then, two TE-polarization scattering problems are solved to demonstrate the performance of the proposed SS-SIE formulation. 

A personal laptop with Intel i7-7770 3.6 GHz CPU and 48G memory is used for all the experiments in this Section. Our in-house codes, including the proposed SS-SIE formulation and the PMCHWT formulation, are implemented in Matlab, and full vectorization is exploited for both approaches to enhance the performance. Only a single thread is used for a fair comparison. {In addition, the Legendre polynomial is used for the general Gaussian quadrature rule to compute the Gaussian points and weights.}
\subsection{Validation of {the Proposed Integration Approach}}
\begin{figure}
	\centerline{\includegraphics[width=4cm]{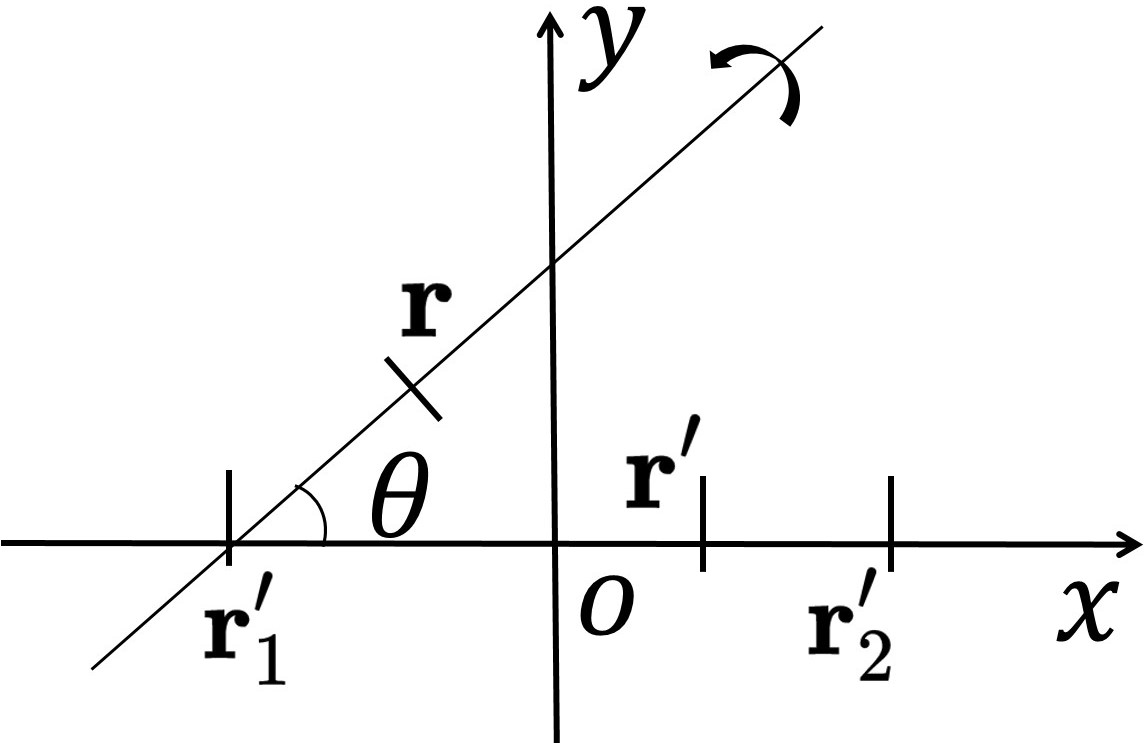}}
	\caption{{A source segment with the endpoints $\mathbf{r}_{1}^{\prime}(-0.065,0,0)\;[\text{m}]$ and $\mathbf{r}_{2}^{\prime}(0.065,0,0)\;[\text{m}]$ is considered. $\mathbf{r}$ is the observation point, which is $0.02$ m away from $\mathbf{r}_{1}^{\prime}$. $\theta$ changes from $0.05\pi$ to $\pi$.}}
	\label{confiforsing}
\end{figure}

{An example presented in Fig. \ref{confiforsing} is used to verify the efficiency and accuracy of the proposed integration approach. As shown in Fig. (\ref{confiforsing}), $\mathbf{r}_{1}^{\prime}(-0.065,0,0)\;[\text{m}]$ and $\mathbf{r}_{2}^{\prime}(0.065,0,0)\;[\text{m}]$ are the endpoints of source segment, and the observation point $\mathbf{r}$ is $0.02$ m away from $\mathbf{r}_{1}^{\prime}$. $\theta$ is the angle between ${(\mathbf{r}_{1}^{\prime}-\mathbf{r})}$ and ${(\mathbf{r}_{1}^{\prime}-\mathbf{r}_{2}^{\prime})}$. (\ref{L}) and (\ref{K}) are used to calculate the integration when $\left|k{(\mathbf{r-r'})}\right| < \delta$, and the Gaussian quadrature is used for the remaining parts. In the computation, we use $k=1$ m$^{-1}$ for convenience. }
{We used the proposed integration approach and the Gaussian quadrature to calculate $\mathcal{L}$ and $\mathcal{K}$ operators. We keep increasing the integration points in the Gaussian quadrature until $|(t_{n+1}-t_n)|/t_{n+1}<10^{-15}$, where $t_n$ denotes the integration results obtained from the Gaussian quadrature with $n$ integration points.} { When the results of $\mathcal{L}$ and $\mathcal{K}$ operators reach convergence precision, we count the number of points used for Gaussian quadrature and the relative error between the two approaches, which represents the overall accuracy of the {proposed integration approach}.}

\begin{figure}[H]
	\begin{minipage}{0.47\linewidth}
		\centerline{\includegraphics[scale=0.055]{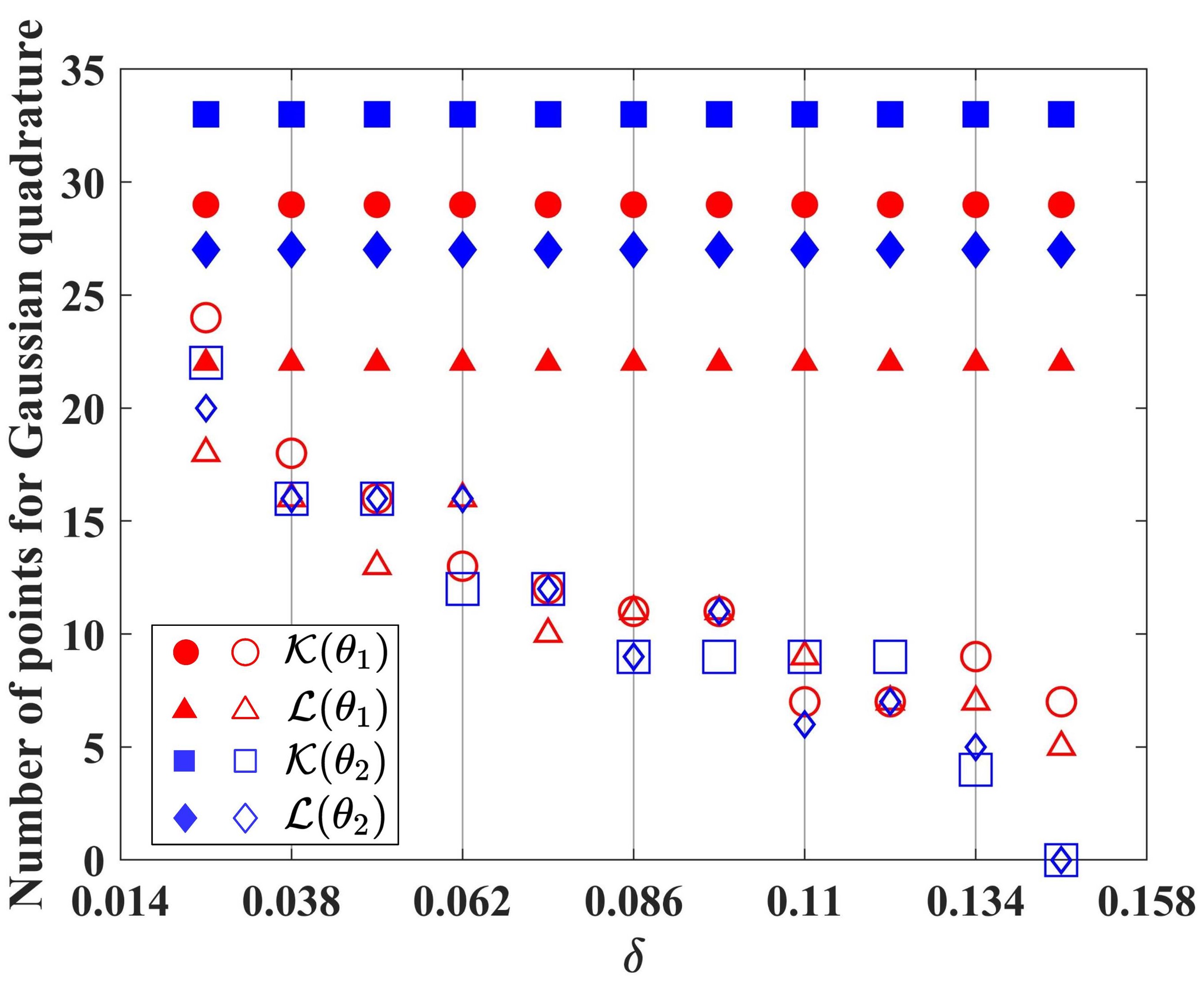}}
		\centerline{(a)}
	\end{minipage}
	\hfill
	\begin{minipage}{0.47\linewidth}
		\centerline{\includegraphics[scale=0.055]{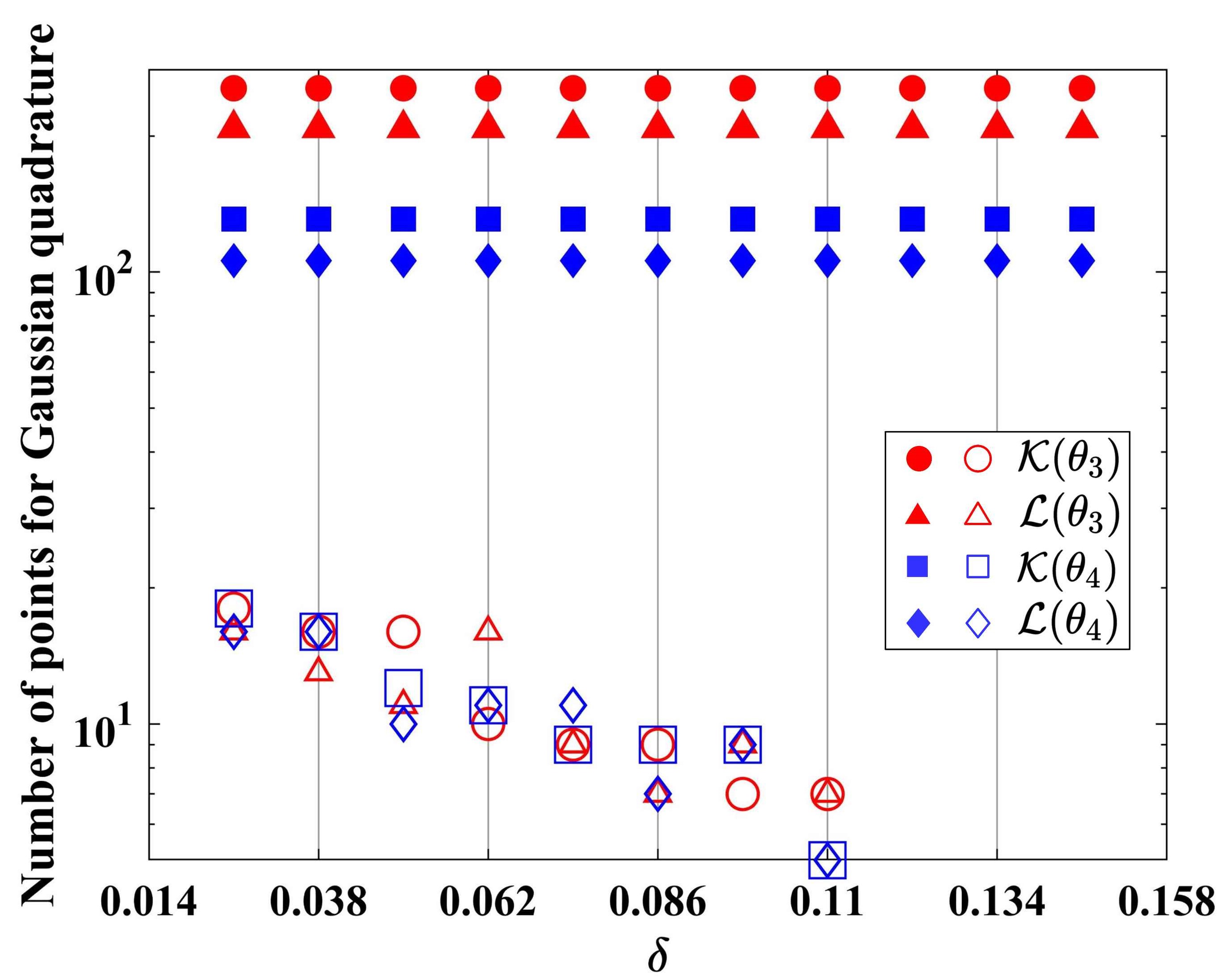}}
		\centerline{(b)}
	\end{minipage}
	\vfill
	\begin{minipage}{0.45\linewidth}
		\centerline{\includegraphics[scale=0.055]{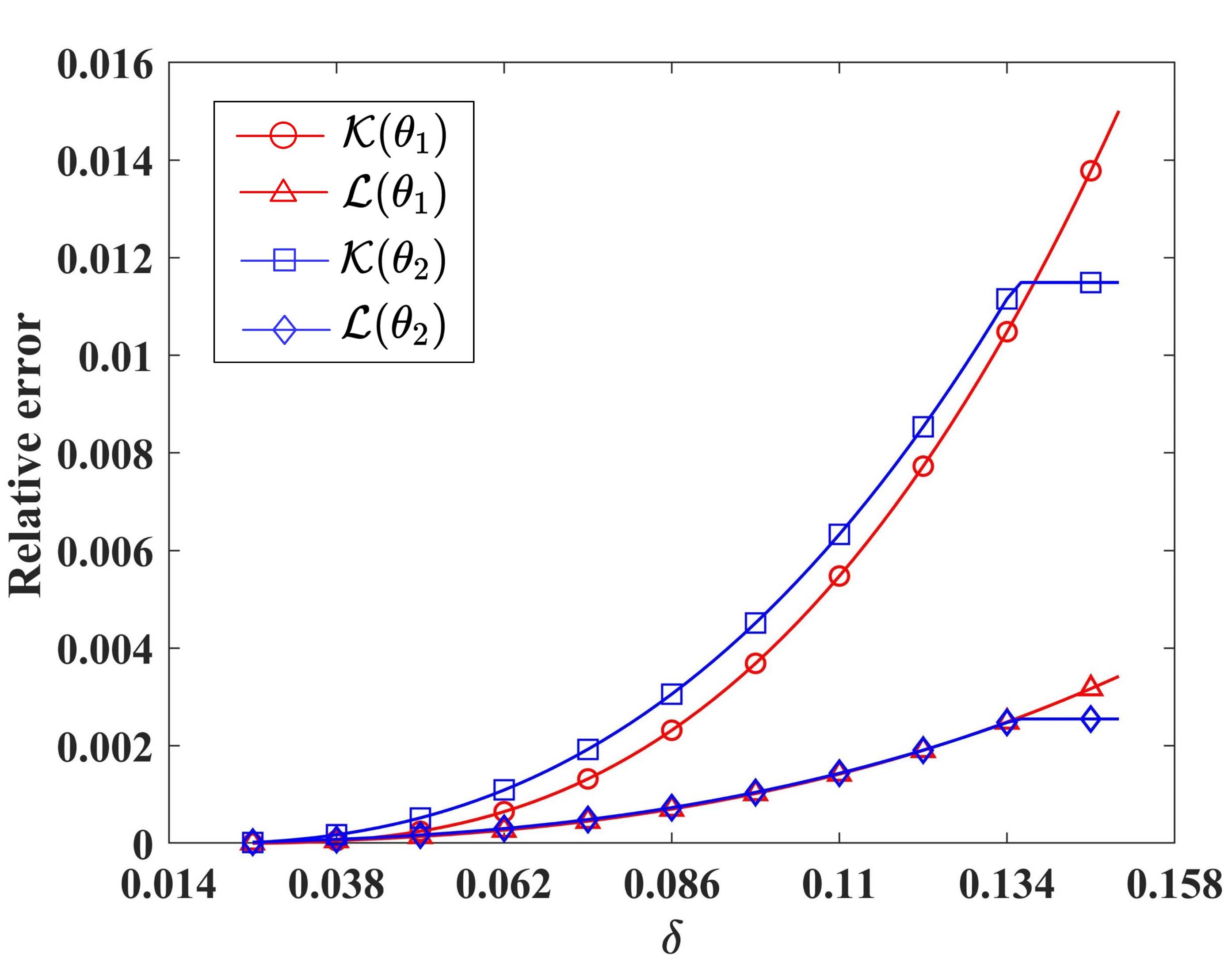}}
		\centerline{(c)}
	\end{minipage}
	\hfill
	\begin{minipage}{0.49\linewidth}
		\centerline{\includegraphics[scale=0.055]{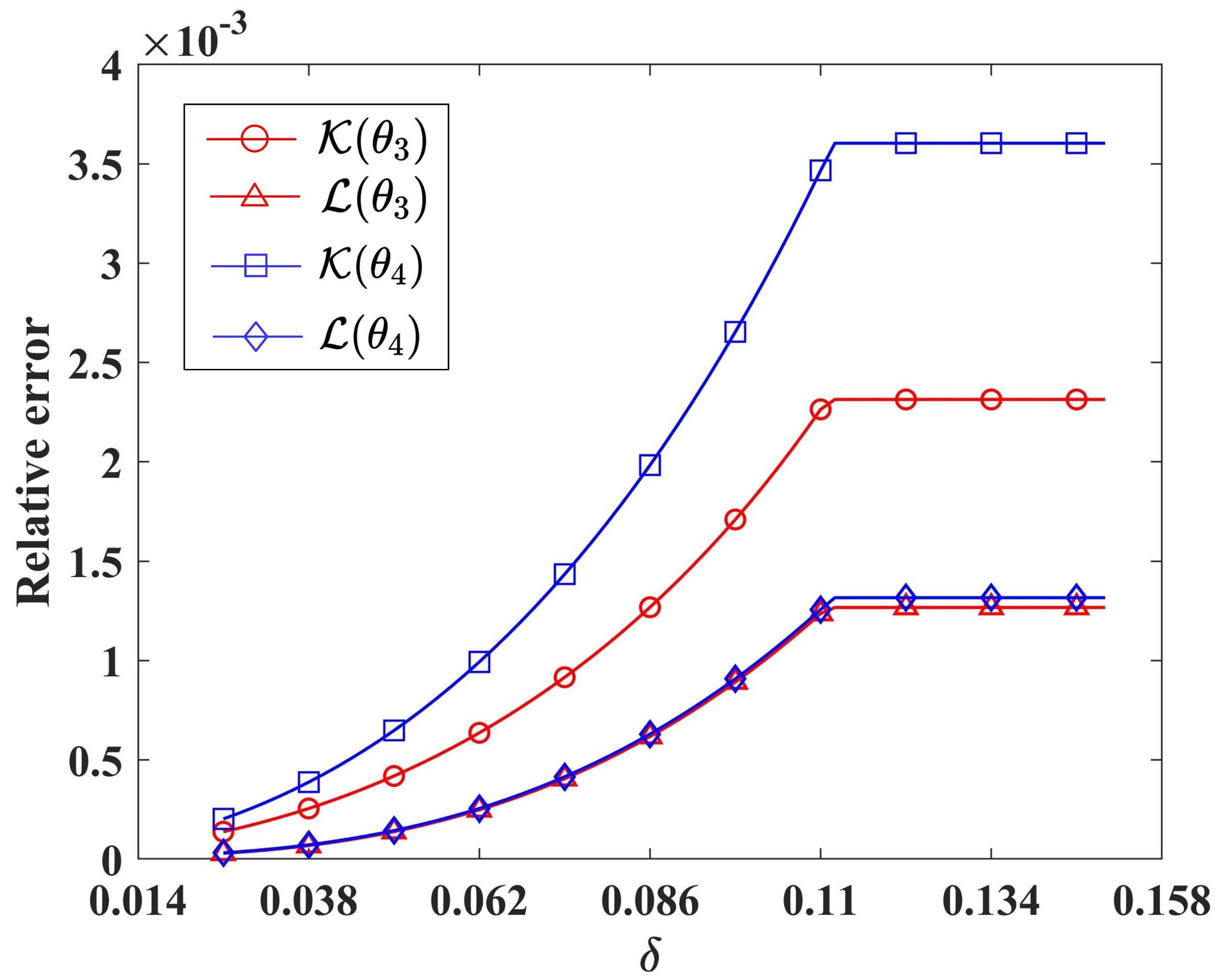}} 
		\centerline{(d)}
	\end{minipage}
	\caption{{(a) (b) The color-filled marks are for the fully Gaussian quadrature approach, and the hollow ones are for the proposed integration approach. $\mathcal{K}(\theta)/\mathcal{L}(\theta)$ means that the results are for the $\mathcal{K}/\mathcal{L}$ operator with a certain $\theta$. (c) (d) The corresponding precision when the results reach convergence. $\theta_{1}=\pi$, $\theta_{2}=0.5\pi$ in (a) and (c). $\theta_{3}=0.05\pi$, $\theta_{4}=0.1\pi$ in (b) and (d).}}
	\label{0609}
\end{figure}

{From Fig. \ref{0609}(a) and (b), it can be found that the {number} of integration points used for Gaussian quadrature increases when $\delta$ decreases. Meanwhile, the overall accuracy of {the proposed integration approach} is higher when $\delta$ decreases, as shown in Fig. \ref{0609}(c) and (d). This is because a small $\delta$ can better approximate the Hankel function and its gradient, as shown in {Fig. \ref{curve}}. Therefore, a balance between computational efficiency and accuracy is needed. In Fig. \ref{0609}(b), the number of integration points required for the Gaussian quadrature is large, since $\theta$ is very small and strong nearly singular integrals are involved in $\mathcal{L}$ and $\mathcal{K}$. Therefore, we need an appropriate $\delta$ to accurately and efficiently calculate the integration. {As shown in Fig. \ref{0609}, $\delta = 0.1$ is accurate enough for most applications. The proposed integration approach only needs less than ten integration points, which is less than half of integration points with the Gaussian quadrature.} It should be noted that the $y$-axis in Fig. \ref{0609}(b) is in log scale for better visualization. It can be found that there are no marks in the large $\delta$ region of Fig. \ref{0609}(b), and the curves of Fig. \ref{0609}(c) and (d) become flat when $\delta$ is larger than specific values. This is because the small variable approximation condition is satisfied in the whole source segment, and (\ref{L}) and (\ref{K}) are used to calculate the integration without the Gaussian quadrature.}

\subsection{A Single Penetrable Object}
\begin{figure}
	\begin{minipage}{0.32\linewidth}
		\centerline{\includegraphics[scale=0.032]{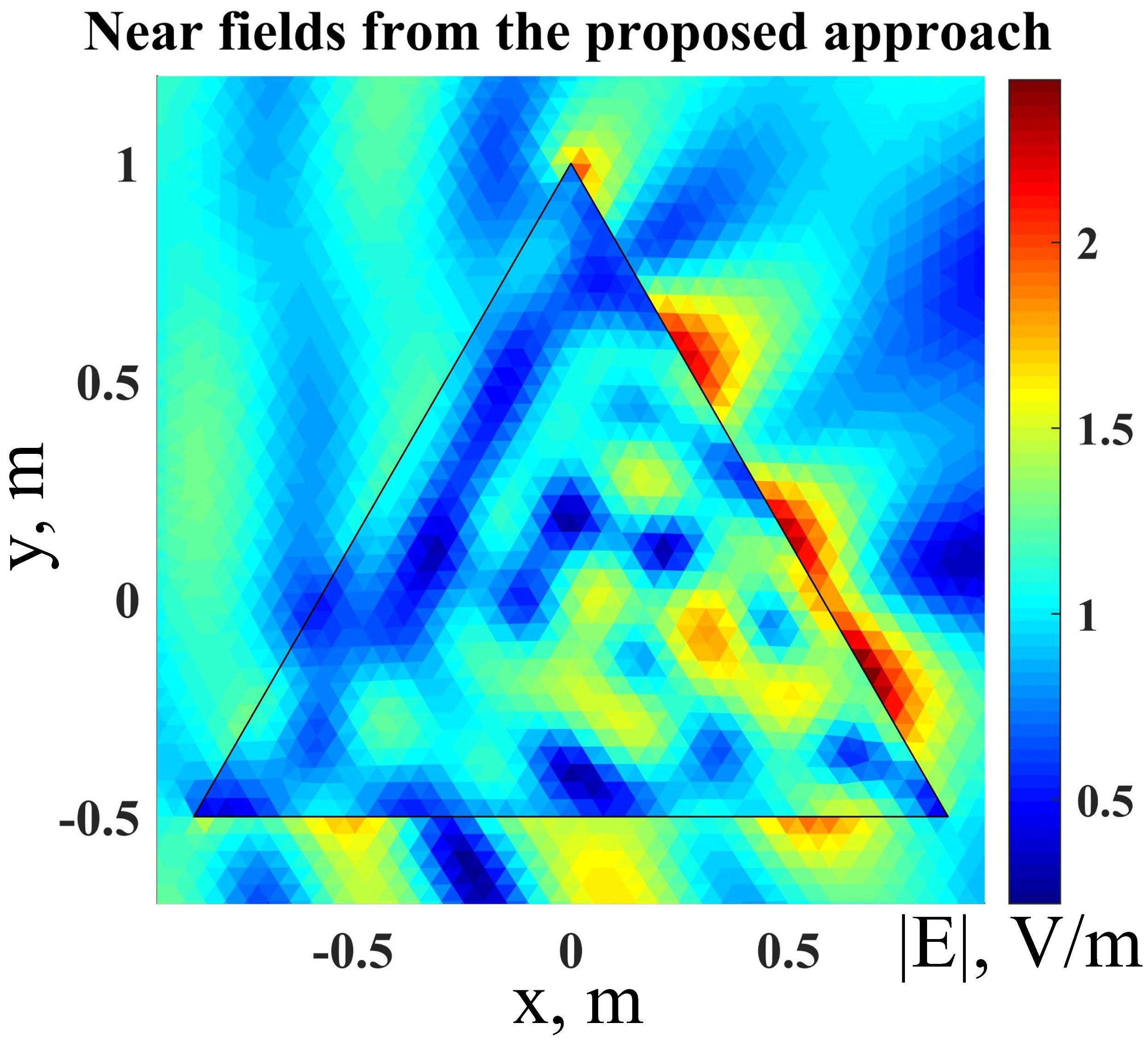}}
		\centerline{(a)}
	\end{minipage}
	\hfill
	\begin{minipage}{0.32\linewidth}
		\centerline{\includegraphics[scale=0.032]{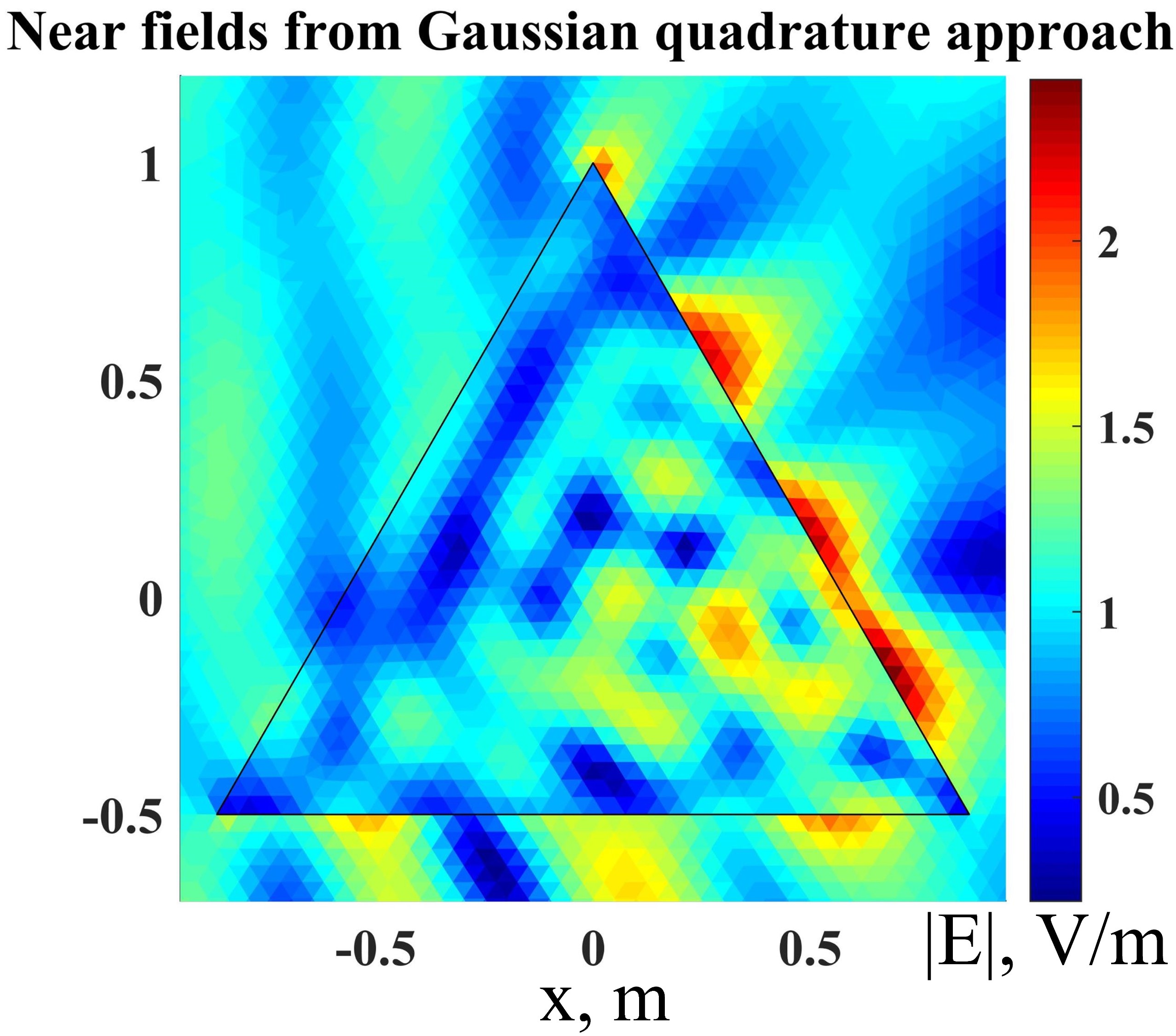}}
		\centerline{(b)}
	\end{minipage}
	\hfill
	\begin{minipage}{0.32\linewidth}
		\centerline{\includegraphics[scale=0.032]{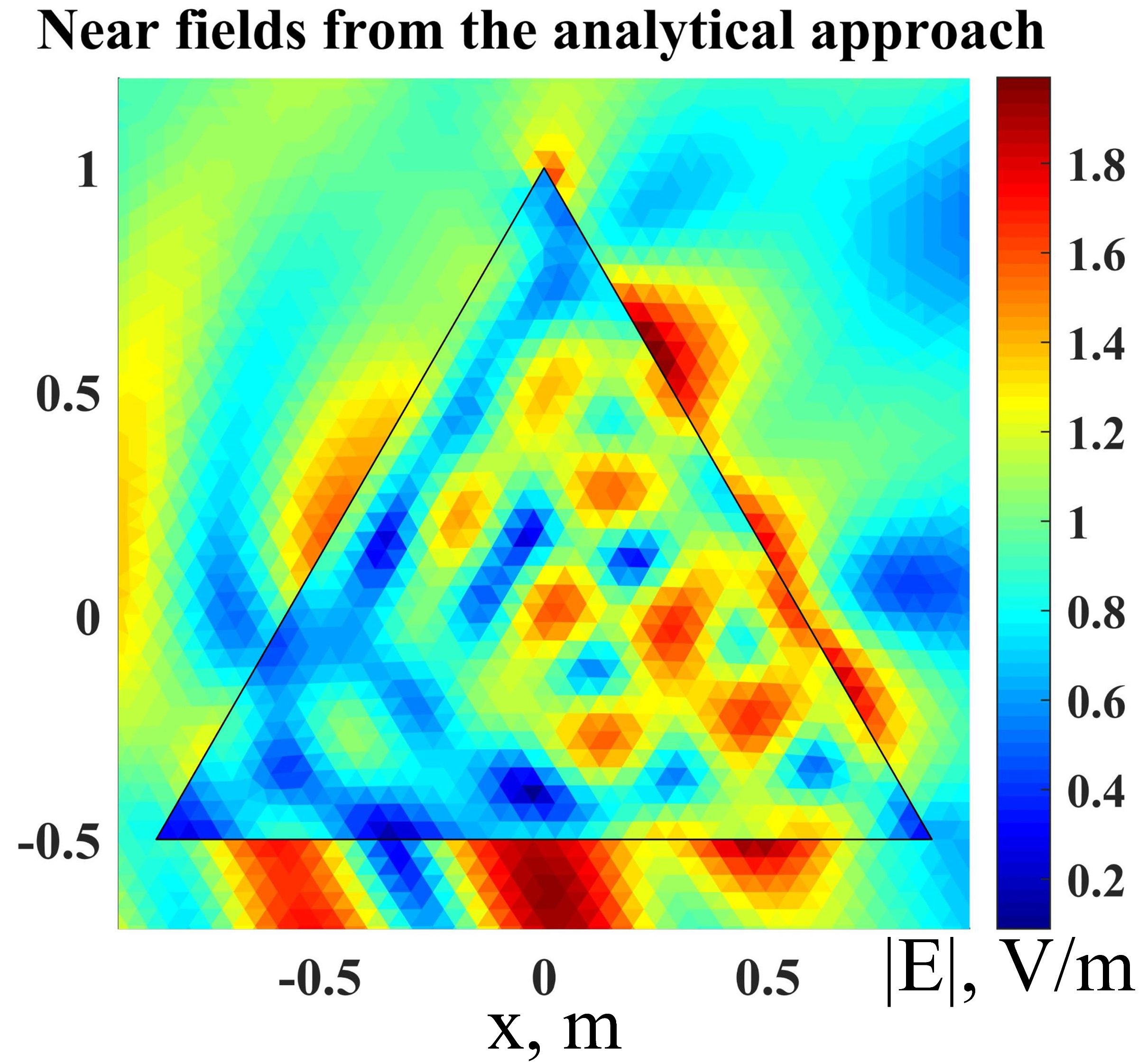}}
		\centerline{(c)}
	\end{minipage} 
	\vfill
	\begin{minipage}{0.32\linewidth}
		\centerline{\includegraphics[scale=0.032]{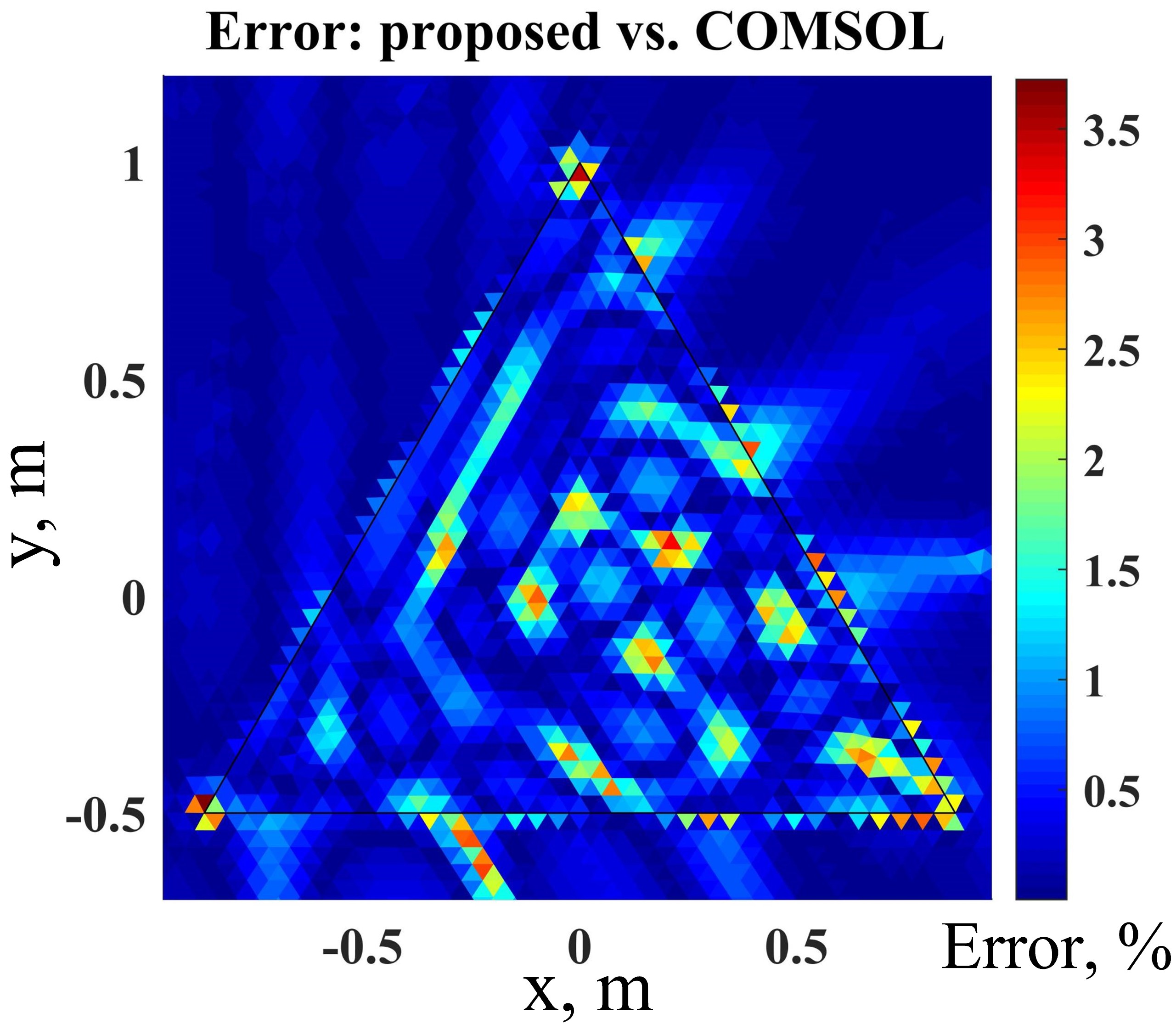}}
		\centerline{(d)}
	\end{minipage}
	\hfill
	\begin{minipage}{0.32\linewidth}
		\centerline{\includegraphics[scale=0.032]{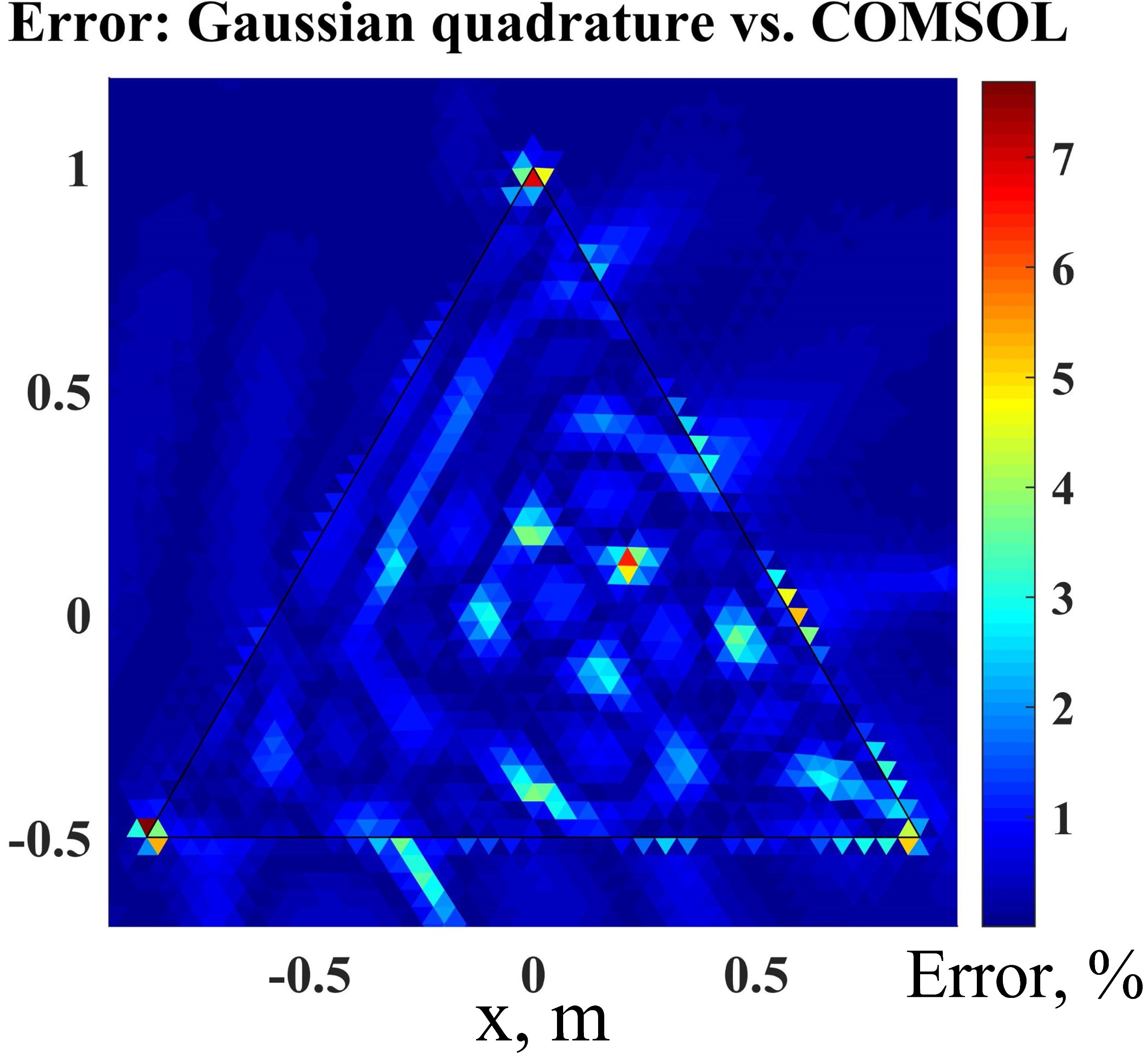}}
		\centerline{(e)}
	\end{minipage}
	\hfill
	\begin{minipage}{0.32\linewidth}
		\centerline{\includegraphics[scale=0.032]{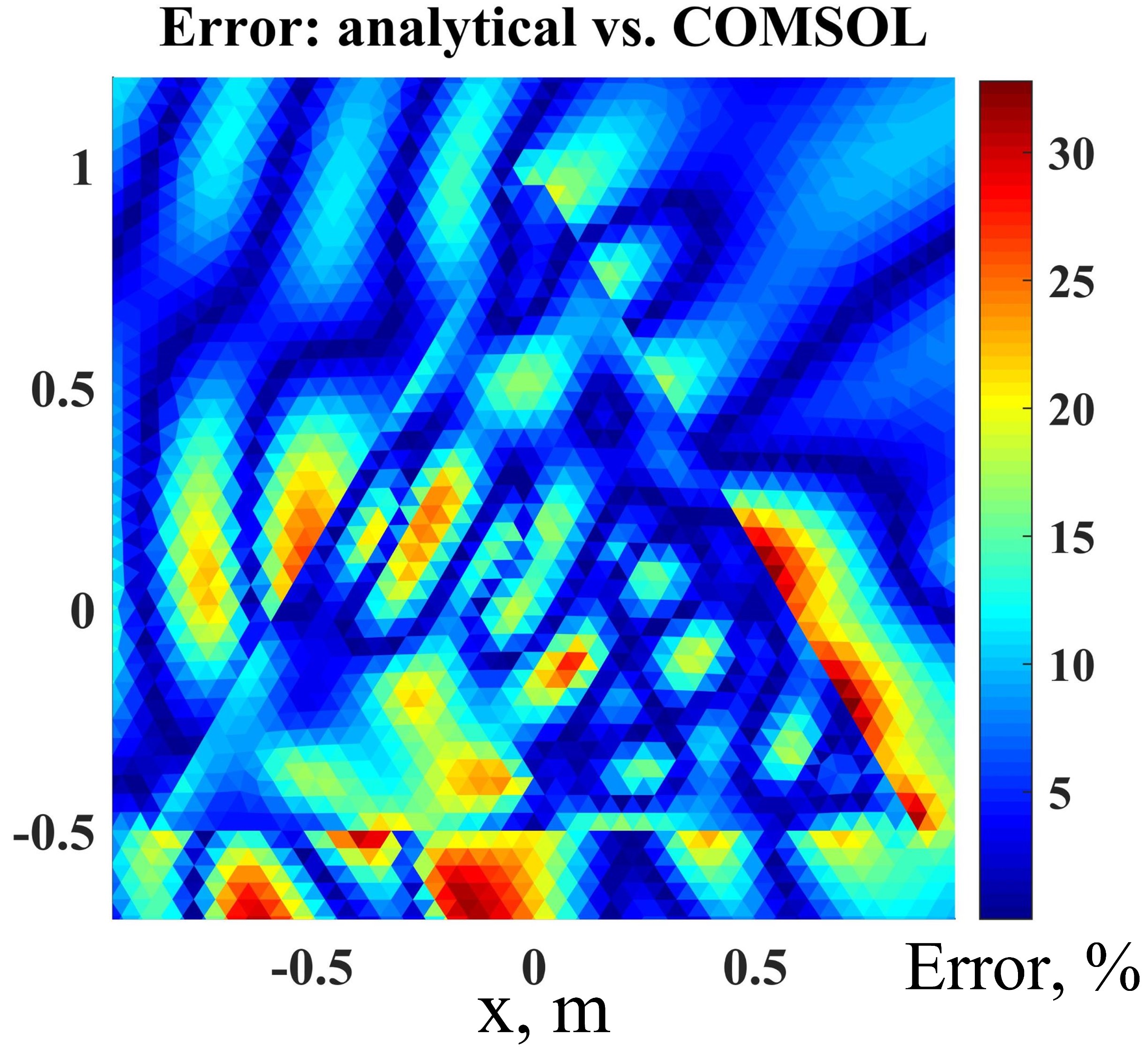}}
		\centerline{(f)}
	\end{minipage}
	\caption{ {The near fields from (a) {the proposed integration approach}, (b) the Gaussian quadrature and (c) the analytical approach. (d) (e) (f) The relative errors compared with the results from COMSOL.}}
	\label{Tri_near}
\end{figure}	
\begin{figure}
	\begin{minipage}{0.55\linewidth}
		\centerline{\includegraphics[scale=0.049]{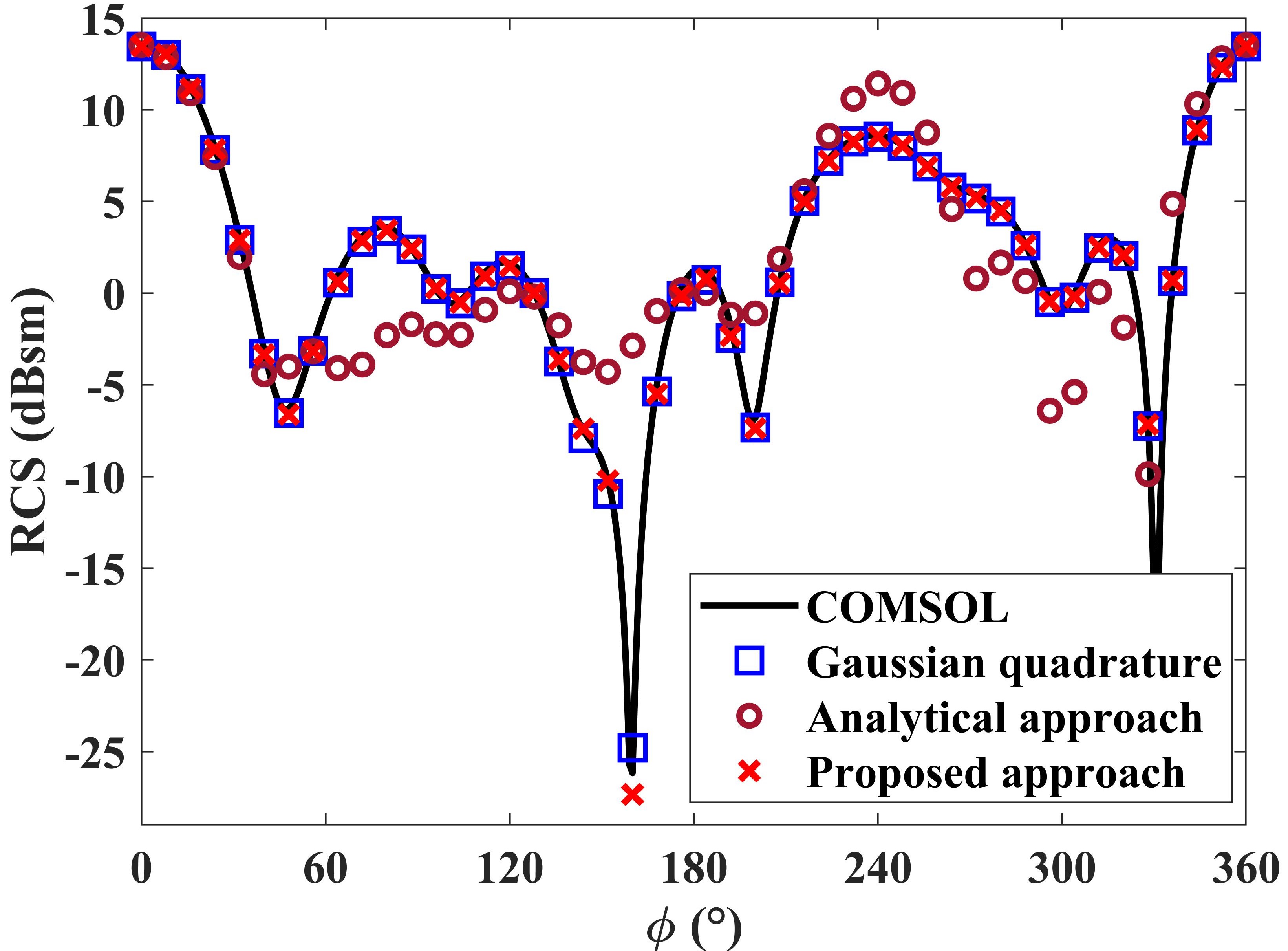}}
		\centerline{(a)}
	\end{minipage}
	\hfill
	\begin{minipage}{0.44\linewidth}
		\centerline{\includegraphics[scale=0.18]{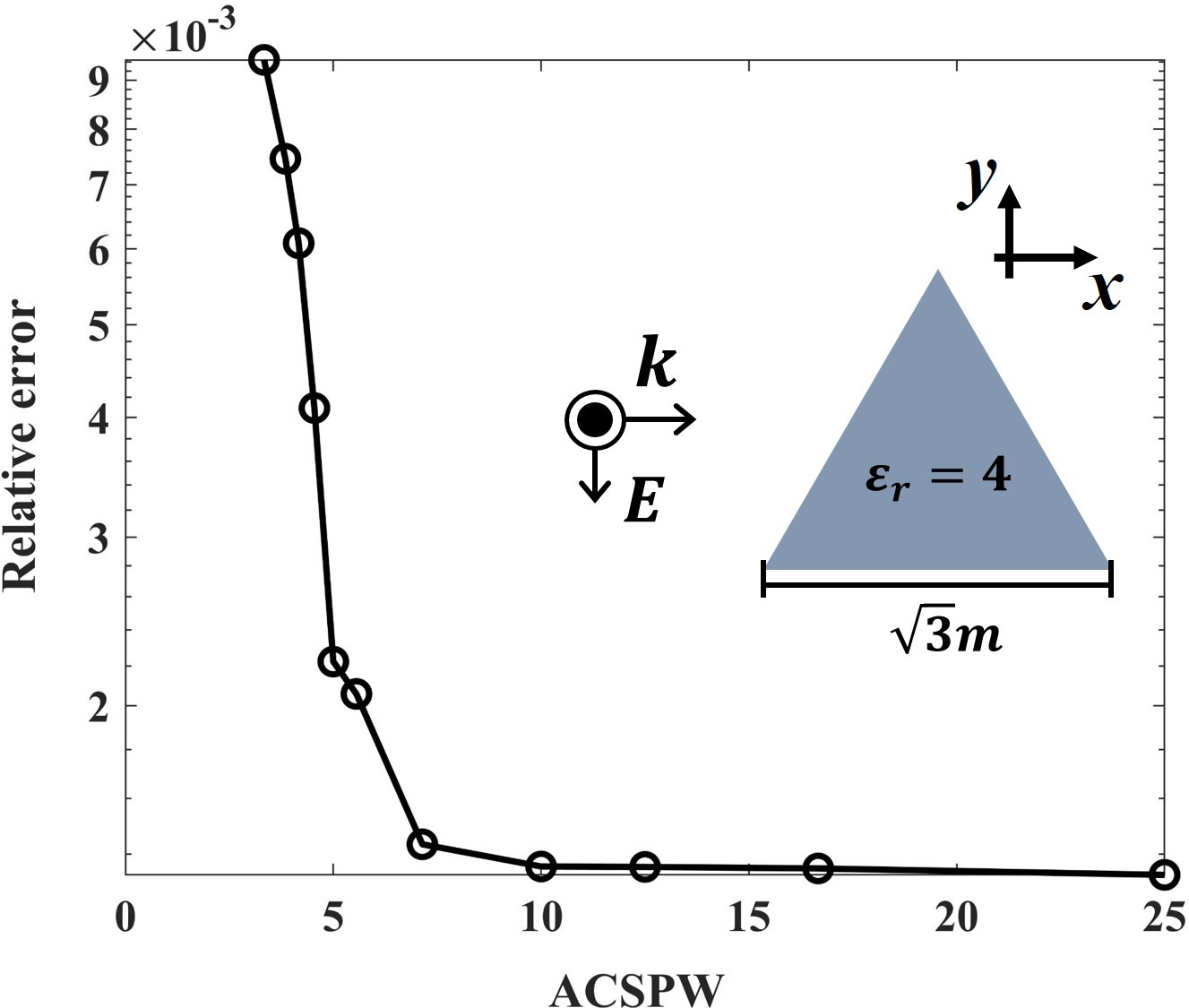}}
		\centerline{(b)}
	\end{minipage}
\caption{(a) RCS obtained from the proposed SS-SIE formulation, the analytical approach, the fully Gaussian quadrature approach, and the COMSOL. (b) The relative error compare to the results from COMSOL when different mesh size is used.}
\label{Tri-RCS}
\end{figure}
An object whose cross-section is an equilateral triangle is carried out to validate the accuracy and efficiency of {the proposed integration approach}. The side length of the triangle is $\sqrt{3}$ m, and the relative permittivity inside it is $4$. The surrounding medium is air. A TE-polarized plane wave with the frequency of $300$ MHz incidents from the $x$-axis. The averaged length of the segments used to discretize the contour of the triangle is $\lambda / 10$, where $\lambda$ is the wavelength inside the penetrable object. In our comparisons, the proposed SS-SIE formulation incorporated with {three integration approaches (namely, {the proposed integration approach}, the analytical approach {[\citen{TraditionalM1}, Ch.3, pp. 43-44]} \cite{TraditionalM2}, and the Gaussian quadrature}), are used. In the analytical approach, only the self segments are treated to be singular. Then, the Green's function and its gradient are approximated by (\ref{SmallAforG}) and (\ref{SmallAforGG}), and analytical integration is applied. Other integration are calculated through Gaussian quadrature. 14-point and 5-point Gaussian quadrature for the source and observation line integration are used. 

The near- and far-fields obtained from the three approaches are shown in Fig. \ref{Tri_near} and \ref{Tri-RCS}. The reference results are calculated from the COMSOL with fine mesh. The relative error of the proposed SS-SIE formulation is less than 4\% compared with the COMSOL, as shown in Fig. \ref{Tri_near}(d). In addition, it is found that the accuracy of {the proposed integration approach} is about twice better than that of the Gaussian quadrature, as shown in Fig. \ref{Tri_near}(d) and (e). The RCS obtained from the proposed integration approach and the Gaussian quadrature shows excellent agreements with the results from the COMSOL. However, the results from the analytical approach are the worst among the three formulations for both near fields and far fields, as shown in Fig. \ref{Tri_near}(f) and Fig. \ref{Tri-RCS}(a). {The reason may be the inconsistency of the accuracy. It can be found from Section V-A that $14$ points may not be accurate enough to evaluate the integration, especially when the observation point is very close to the source segment (e.g., at vertices of the triangle). In the analytical approach, the results are accurate when the observation segment is overlapped with the source segment, but are not accurate enough when the observation segment is close to the source segment. However, when the Gaussian quadrature is used, the results are much more accurate than those of the analytical approach. This may be due to the fact that the accuracy of the Gaussian integral is uniform when the obesvation segment is overlapped or close to the source segment.}

In Fig. \ref{Tri-RCS}(b), the relative error with respect to averaged count of segments per wavelength (ACSPW) in the free space is shown, and the stability of {the proposed integration approach} is proved again. The {relative} error used here is defined as 
\begin{equation}\label{RE}
	\frac{\sum{_i \left \lVert \text{RCS}^{\text{cal}} \left( \phi _i \right) - \text{RCS}^{\text{ref}}\left( \phi _i \right) \right \rVert}^2}{\sum{_i \left \lVert \text{RCS}^{\text{ref}}\left( \phi _i \right) \right \rVert ^2}},
\end{equation}
where $ \text{RCS}^{\text{cal}} \left( \phi _i \right) $ denotes results calculated from the proposed SS-SIE formulation and $ \text{RCS}^{\text{ref}} \left( \phi _i \right) $ is the reference results obtained from the COMSOL.
\subsection{An Array with $5\times5$ Multilayered Coated Dielectric Cylinders}
\begin{figure}[H]
	\centerline{\includegraphics[scale=0.15]{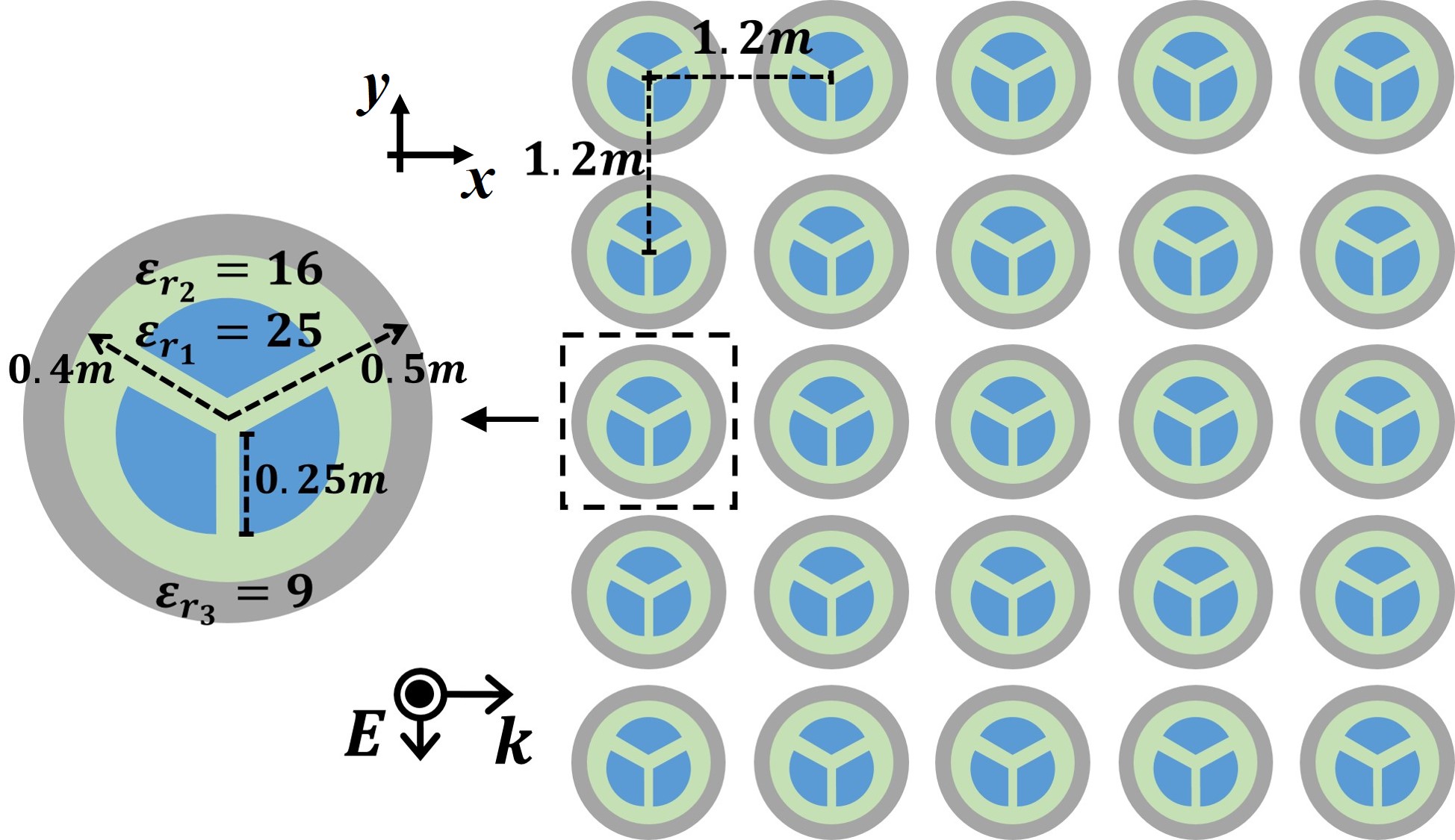}}
	\caption{{An array with 5$\times$5 multi-layered coated dielectric cylinders.}}
	\label{ST_Array}
\end{figure}
{In this subsection, an array with $5\times5$ multilayered coated dielectric cylinder is simulated, as shown in Fig. \ref{ST_Array}. The innermost sectors are dielectric objects with $r_1=0.25$ m and $\varepsilon_{r_1} = 25$. The second layer is dielectric with $r_2=0.4$ m and $\varepsilon_{r_2} = 16$. The third layer is a dielectric ring with $r_3=0.5$ m and $\varepsilon_{r_3} = 9$. The central distance between two adjacent elements is $d=1.2$ m, and the background medium is air. The TE-polarized plane wave with $f=300$ MHz incidents from the $x$-axis. $\lambda_{0}/40$ is used to discretize all boundaries, where $\lambda_{0}$ is the free-space wave length.}

%
\begin{figure}[H]
	\begin{minipage}{0.49\linewidth}
		\centerline{\includegraphics[scale=0.05]{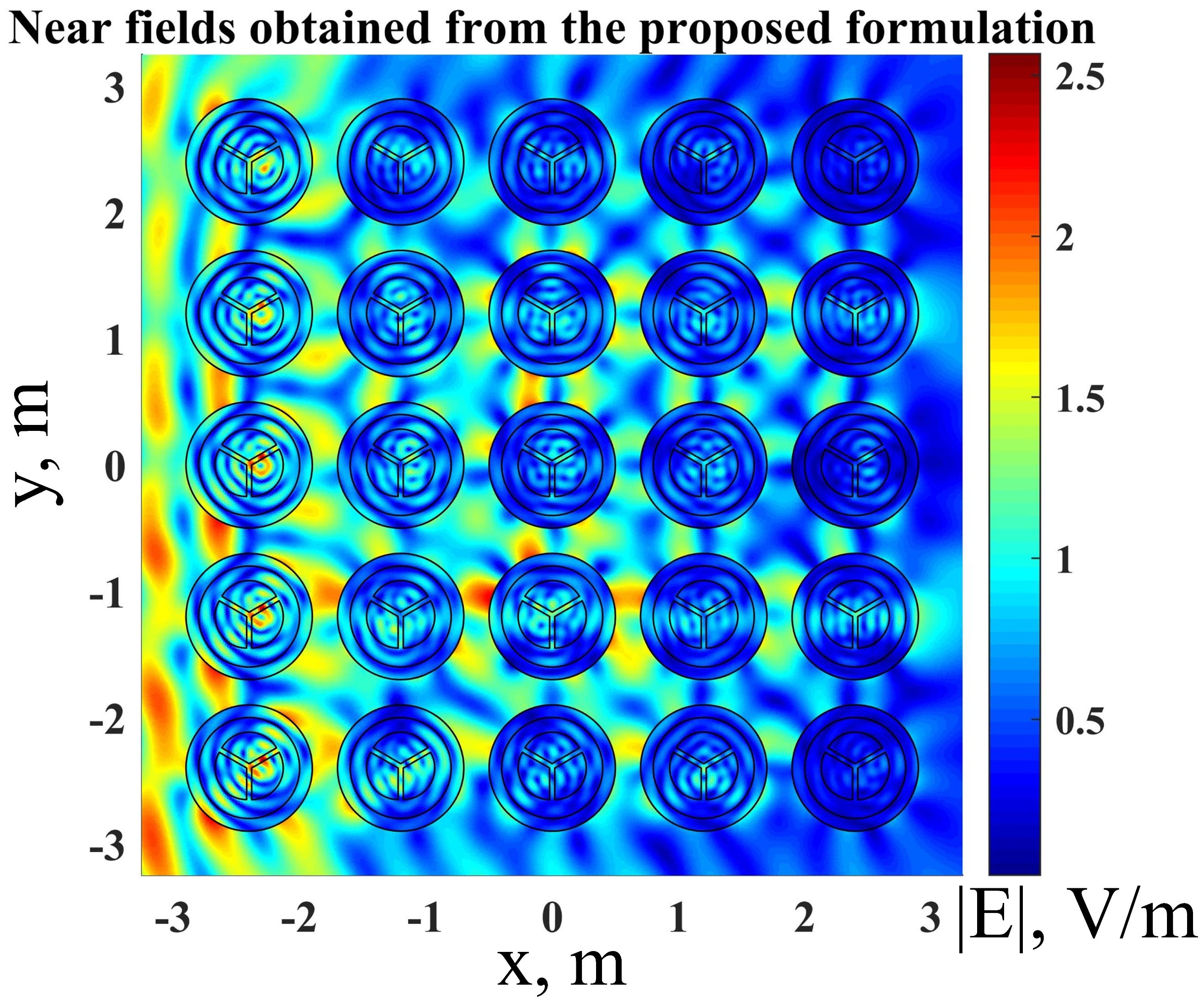}}
		\centerline{(a)}
	\end{minipage}
	\hfill
	\begin{minipage}{0.49\linewidth}
		\centerline{\includegraphics[scale=0.05]{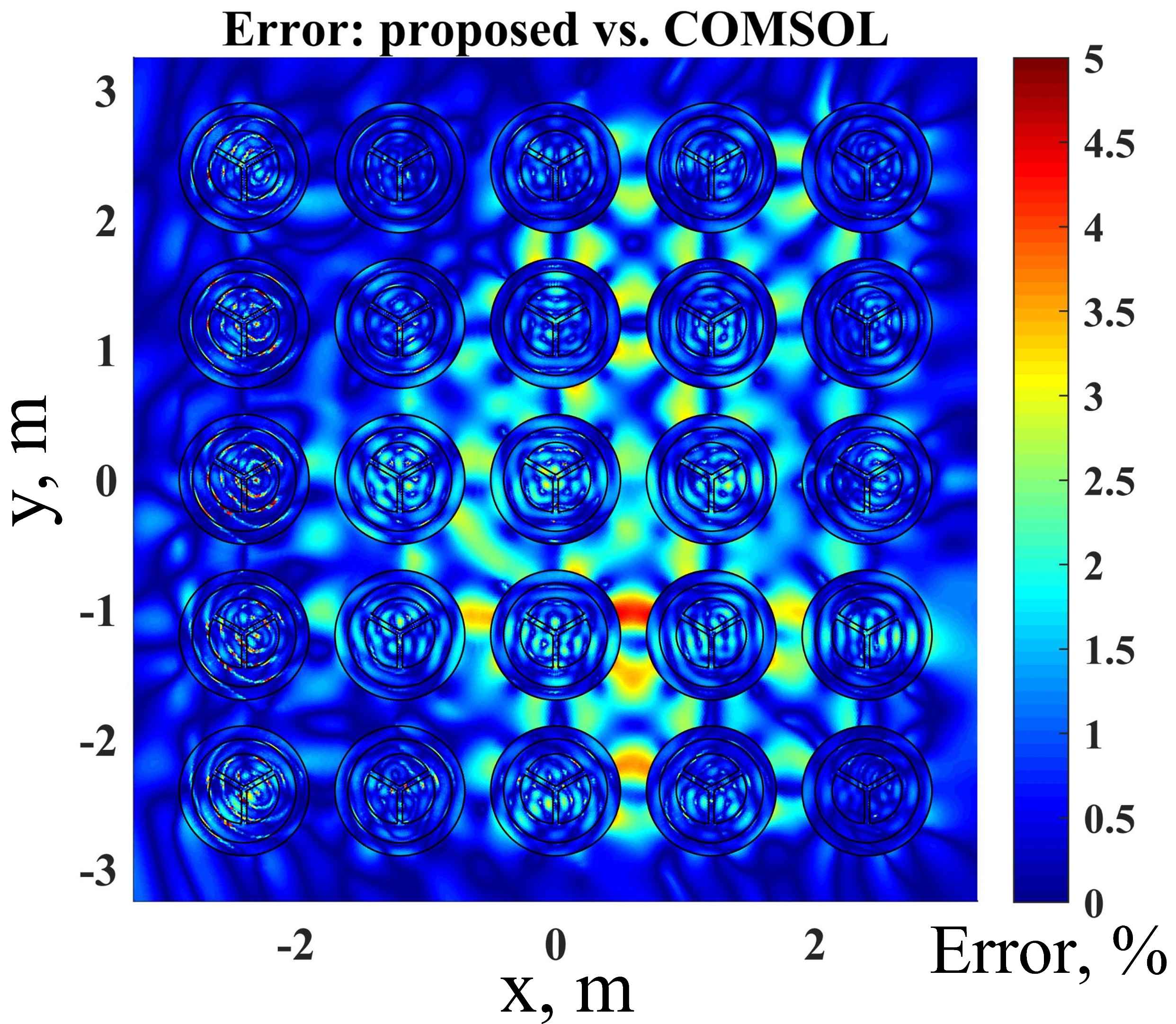}}
		\centerline{(b)}
	\end{minipage} 
	\caption{{(a) The near field from the proposed SS-SIE formulation for the array in Fig. \ref{ST_Array}. (b) The relative error compared with COMSOL.}}
	\label{array_near}
\end{figure}

\renewcommand\arraystretch{1.5}
\begin{table}[H]
	\centering
	\caption{Comparison of computational cost for the PMCHWT formulation and the proposed SS-SIE formulation}\label{table2}
	\begin{tabular}{llll}
		\hline
		\hline
		&\textbf{PMCHWT}&\textbf{Proposed} &\textbf{Ratio}\cr  
		\hline
		\hline
		\textbf{Overall Time Cost} [s]             & 3,600            & 1,037   &    0.29   \\
		\hline
		Time for Matrices Filling [s]               &   3420          & 1,039    &   0.30   \\
		\hline
		Time for $\mathbb{Y}_s$ Generation[s]               &  -----          & 0.4    & -----     \\
		\hline
		Time for Equation Solving [s]               &    176        & 5.6       &  0.03  \cr
		\hline
		\textbf{Memory Consumption} [MB]                &    5189             & 1,331   &   0.26    \\
		\hline
		\textbf{{Number of Unknowns}}                 &   20,700            & 4,000    &    0.19   \\
		\hline
		\hline 
	\end{tabular}
\end{table}

\newcommand{\tabincell}[2]{\begin{tabular}{@{}#1@{}}#2\end{tabular}}  
\begin{table}[H]		
	\begin{center}		
		\caption{Comparison of the condition number between the proposed SS-SIE formulation and the PMCHWT formulation}\label{Cond_detial2}
		\tabcolsep 0.05in
		\begin{tabular}{c|c|c|c|c|c|c|c}
			\hline
			\hline
			\multicolumn{8}{c}{Condition Number}\\
			\hline
			\hline
			\multicolumn{7}{c|}{Proposed} &{PMCHWT}   \\
			\cline{1-8}
			\multirow {2}{*}{1st} & $\mathbb{L}^{(1)}_{(1,1)}$ &17 & $\mathbb{L}^{(1)}_{(2,2)}$ &17&$\mathbb{L}^{(1)}_{(3,3)}$ &17 &\multirow {6}{*}{\tabincell{c}{185,017 \\ {(w/o equilibration)}\\ {66,101}\\{(diagonal)}\\ {1,924}\\{(extended diagonal)} }}\\
			\cline{2-7}
			&$\widehat{\mathbb{L}}^{(1)}_{(1,1)}$ &11 & $\widehat{\mathbb{L}}^{(1)}_{(2,2)}$ &11&$\widehat{\mathbb{L}}^{(1)}_{(3,3)}$ &11 \\
			\cline{1-7}
			{2nd}&${\mathbb{C}}_{\gamma_1}$ &161 & ${\mathbb{V}}_{\gamma_4}$ &51&$\widehat{\mathbb{L}}^{(2)}_{(4,4)}$ &101\\
			\cline{1-7}
			{3rd}&${\mathbb{C}}_{\gamma_2}$ &551 & ${\mathbb{V}}_{\gamma_5}$ &131&$\widehat{\mathbb{L}}^{(3)}_{(5,5)}$ &88\\
			\cline{1-7}
			Solving&\multicolumn{3}{c|}{$\mathbb{U} -\widehat{\mathbb{L}} {\mathbb{Y}_{s}}$}&\multicolumn{3}{c|}{\tabincell{c}{1,257\\ {(w/o equilibration)} }}\\
			\hline
			\hline		
		\end{tabular}	
	\end{center}
\end{table}

\begin{figure}[H]
	\begin{minipage}{0.49\linewidth}
		\centerline{\includegraphics[scale=0.05]{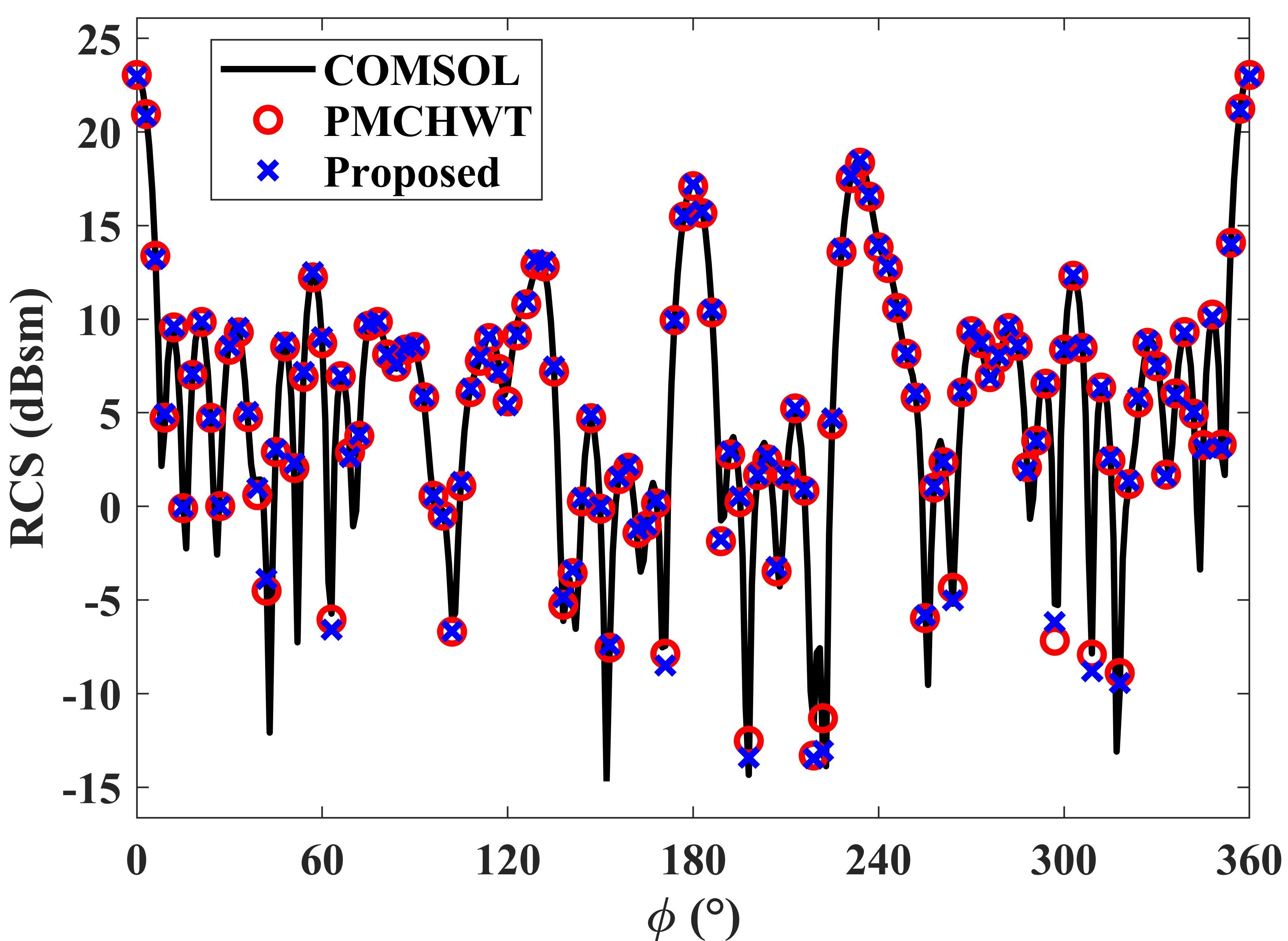}}
		\centerline{(a)}
	\end{minipage}
	\hfill
	\begin{minipage}{0.49\linewidth}
		\centerline{\includegraphics[scale=0.05]{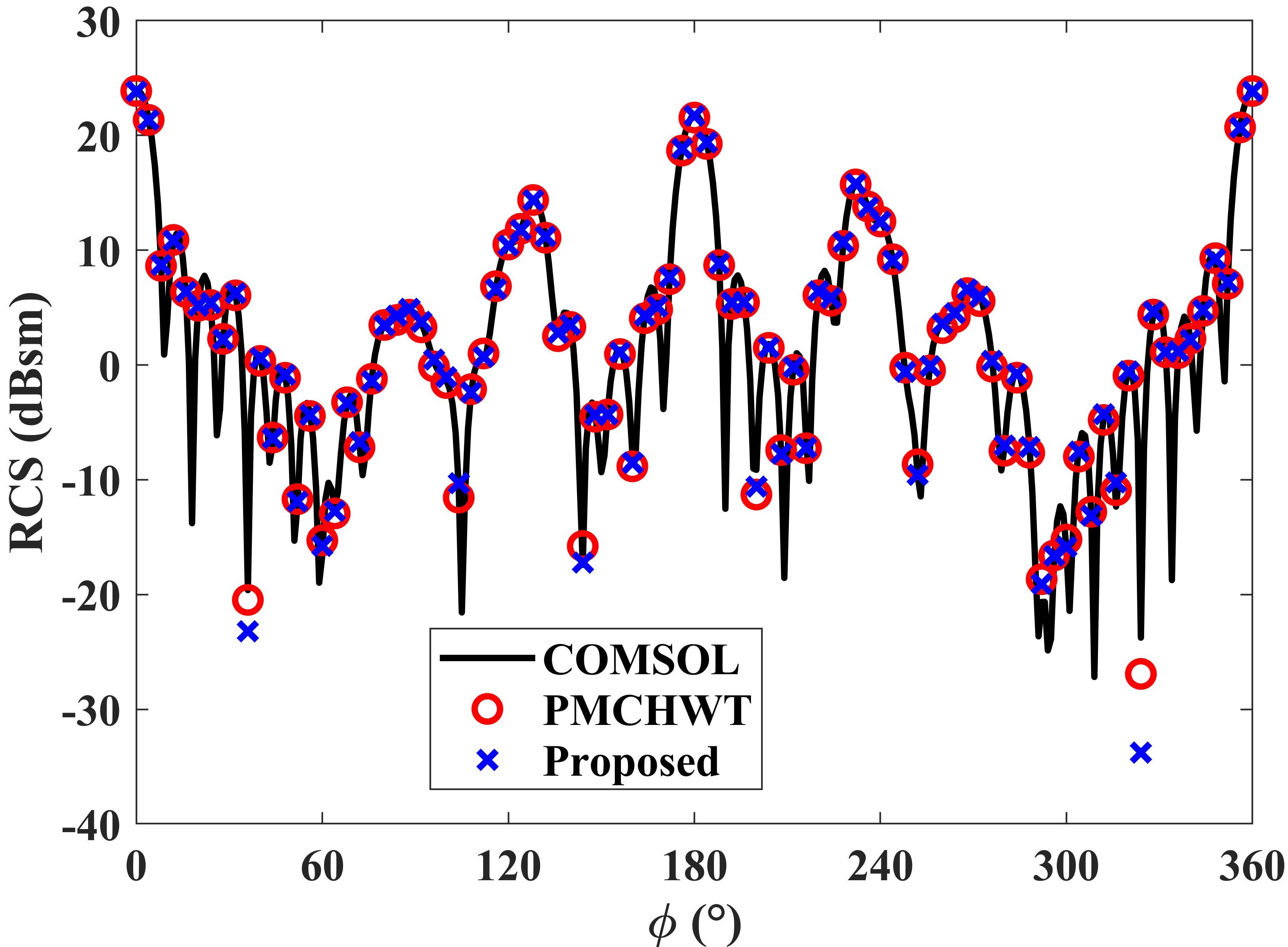}}
		\centerline{(b)}
	\end{minipage} 
	\caption{{ RCS obtained from the COMSOL, the PMCHWT formulation, and the proposed SS-SIE formulation in case of (a) low permittivity and (b)} {high permittivity}.}
	\label{Array_RCS}
\end{figure}

{The near fields of the array are shown in Fig. \ref{array_near}(a),  and the relative error compared with the COMSOL is presented in Fig. \ref{array_near}(b). It can be found that the relative error is less than $5\%$. The RCS is computed by the proposed SS-SIE formulation, the PMCHWT formulation, and the COMSOL, as shown in Fig. \ref{Array_RCS}(a). The results obtained from the proposed SS-SIE formulation are in excellent agreement with the results from the PMCHWT formulation and the COMSOL.} 

Table \ref{table2} shows the computational cost of the PMCHWT formulation and the proposed SS-SIE formulation. PMCHWT formulation has {$20,700$} unknowns while the proposed SS-SIE formulation only requires {$4,000$}. This smaller number is due to the fact that the only electric current density residing on the outermost boundary is required with the proposed SS-SIE formulation. For the memory consumption, {$5,189$} MB memory is used in the PMCHWT formulation, only {$1,331$} MB memory, $26\%$ of the PMCHWT formulation,  is used with the proposed SS-SIE formulation. The CPU time to fill the coefficient matrix is {$3,420$} s with the PMCHWT formulation and {$1,039$} s with the proposed SS-SIE formulation. Therefore, the CPU time is reduced by {$71\%$}. {Since there is only one type of scatters in this array, the DSAO is needed to be calculated once, which uses $0.4$ s}. Therefore, the saving in terms of memory and CPU time is significant with the proposed formulation. {Although there are 13 matrices to be inverted with the proposed formulation, all the matrices have smaller dimensions with the largest dimension being 4,000.}

To construct the surface equivalent current density on the outermost boundary, the surface equivalence theorem is applied three times. As shown in Table \ref{Cond_detial2}, the matrix condition number of the PMCHWT formulation is {$185,017$}. However, the condition number of the final matrix of the proposed approach is only {$1,257$}. {For the PMCHWT formulation, all geometric details are in the final large matrix, but they are implicitly incorporated into several small matrices with the proposed SS-SIE formulation.} Therefore, the condition number of the final system is smaller with the proposed SS-SIE formulation.

 {There are several methods to decrease the condition number in SIE, like matrix equilibration [\citen{MatrixEq}]. We applied diagonal scaling and extended scaling to balance the source and functions in both the PMCHWT formulation and the proposed formulation. As shown in Table \ref{table2}, the condition number decreases significantly in PMCHWT formulation. However, the matrix equilibration has little effect on the proposed formulation. This may be due to the fact that only a single electric current and EFIE are applied in the proposed formulation.}

{In addition, the same structure with high permittivity, which is $\varepsilon_{r_1} = 125$ for the innermost objects, $\varepsilon_{r_2} = 75$ for the second layer, and $\varepsilon_{r_3} = 25$ for the third layer, is modeled to validate the robustness of the proposed SS-SIE formulation. Fig. \ref{Array_RCS}(b) shows the RCS with the the PMCHWT formulation, the proposed formulation, and the COMSOL. They show excellent agreement with each other. Therefore, the proposed SS-SIE formulation can model the structures with high permittivity.}

\subsection{Discussion}
{Assuming there are $p$ layers with $N_1, N_2, N_3,\cdot\cdot\cdot, N_p$ segments on each boundary, and $N=N_1+N_2+…+N_p$ segments in all. For the matrix filling, the computational complexity of the proposed SS-SIE formulation is more than $O(N_1^2)+ O(N_2^2)+….+ O(N_p^2)$ and less than $O(N^2)$. For the matrix equation solving, the computational complexity is $O(N_1^3)+O(N_2^3)+…+O(N_{i-1}^3)+O(N_i^3)$. Therefore, the proposed formulation has a lesser computational complexity than that of the PMCHWT formulation in most occasions ($O(N^2)$ for the matrix filling and $O(N^3)$ for matrix solving).}

{However, the proposed SS-SIE formulation suffers from efficiency issue for large structures since the construction of the DSAO requires dense matrix inversion. The overhead cannot be ignored for those large structures. There are basically two ways to overcome this issue. One way is to use efficient direct algorithms, such as H-Matrix [\citen{HMATRIX}], HSS solver [\citen{HSS}], and so on. The other efficient way is to partition the large objects into small ones. The DSAO is constructed for those small objects, and the efficiency is improved.} 

{In addition, the proposed formulation may have problems at the resonant frequencies since only the EFIE formulation is used to solve the scattering problems. At the resonance, wrong current and large condition number may be obtained due to mode degeneration. In [\citen{DSAOARBSHAPEDO}], a differential surface admittance operator is defined and combined integral field equations are used to avoid resonance.}

\section{Conclusion}
The surface equivalence theorem states that a closed surface with the appropriate surface equivalent current sources can generate external fields precisely the same as those in the original problems. In this paper, we thoroughly explore this possibility with the method of moments implementations. A novel vector SS-SIE is proposed to solve TE scattering problems for objects embedded in {cylindrical multilayers}. The proposed method employs the full vector electromagnetic formulations and extends our previous work in \cite{MYARXiv} which only works for the scalar TM-polarized electromagnetic problems. 

The proposed SS-SIE formulation only needs a single electric current density on the outermost boundary. It is derived by recursively applying the surface equivalence theorem on each boundary from inner to exterior regions. Compared with the PMCHWT formulation, the proposed SS-SIE formulation can improve computational efficiency in terms of {the number} of unknowns, memory consumption, CPU time, and the system's conditioning number. In addition, to accurately handle the nearly singular and singular integrals in the proposed SS-SIE formulation, {an integration approach} is proposed. Numerical results validate the efficiency and convergence property of {the proposed integration approach} and {SS-SIE formulation}.

Extension of the current work into the three-dimensional case and fast direct solver to accelerate construction ${\mathbb{Y}}_n$ is in progress. We will report more results on this topic in the future.

\section*{Appendix}
According to {the definitions in Section IV}, the following eight identities \cite{Integral1} [\citen{Integral2}, Ch. 5] are used.
\begin{align}\label{I1}
	{I_1} &= \int_{l} {\ln \left| {\mathbf{r}}-{\mathbf{r'}} \right|}  d{\mathbf{r'}} \notag \\[0.2em] 
	& ={l_2 }\ln \left| \mathbf{r}-\mathbf{r}_2 \right| - {l_1}\ln \left| \mathbf{r}-\mathbf{r}_1 \right| 
	 + \left| \mathbf{r}-\mathbf{p} \right|\cdot \notag\\[0.2em] 
	&\ \ \  \left( {{{\tan }^{ - 1}}\frac{{{l_2 }}}{{\left| \mathbf{r}-\mathbf{p} \right|}} - {{\tan }^{ - 1}}\frac{{{l_1}}}{{\left| \mathbf{r}-\mathbf{p} \right|}}} \right) - \left( {{l_2 } - {l_1}} \right), \\[0.5em] 
	{I_2} &= \int_{l} {\vec l \cdot \ln \left| {\mathbf{r}}-{\mathbf{r'}} \right|} {\kern 1pt} {\kern 1pt} {\kern 1pt} d{\mathbf{r'}}{\rm{ = }}\frac{{\vec \tau }}{2} \cdot \{  - \frac{1}{2}[{({l_2 })^2} - {({l_1})^2}] \notag \\[0.2em] 
	& + {\left| \mathbf{r}-\mathbf{r}_2 \right|^2}\ln \left| \mathbf{r}-\mathbf{r}_2 \right| - {\left|  \mathbf{r}-\mathbf{r}_1 \right|^2}\ln \left|  \mathbf{r}-\mathbf{r}_1 \right|\}, \\[0.5em]
	{I_3} &= \int_{l} {\frac{{\rm{1}}}{{{{\left| {\mathbf{r}}-{\mathbf{r'}} \right|}^{\rm{2}}}}}} {\kern 1pt} {\kern 1pt} {\kern 1pt} d{\mathbf{r'}}\\
	&=\frac{{\rm{1}}}{{\left| \mathbf{r}-\mathbf{p} \right|}}{\rm{(}}{\tan ^{ - 1}}\frac{{{l_2 }}}{{\left| \mathbf{r}-\mathbf{p} \right|}}{\rm{  -  }}{\tan ^{ - 1}}\frac{{{l_1}}}{{\left| \mathbf{r}-\mathbf{p} \right|}}{\rm{)}},\\[0.5em] 
	{I_{\rm{4}}} &= \int_{l_i} {\frac{{|\vec l\ |}}{{{{\left| {\mathbf{r}}-{\mathbf{r'}} \right|}^{\rm{2}}}}}} {\kern 1pt} {\kern 1pt} {\kern 1pt} d{\mathbf{r'}}{\rm{ = }} {\rm{(}}\ln \left| \mathbf{r}-\mathbf{r}_2 \right| - \ln \left|  \mathbf{r}-\mathbf{r}_1 \right|),\\[0.5em] 
	{I_5} &= \int_{l} {\frac{{\vec l \cdot \vec l}}{{{{\left| {\mathbf{r}}-{\mathbf{r'}} \right|}^{\rm{2}}}}}} {\kern 1pt} {\kern 1pt} {\kern 1pt} d{\mathbf{r'}}{\rm{ = }}({l_2 } - {l_1}) - {\left| \mathbf{r}-\mathbf{p} \right|^2} \cdot {I_3},\\[0.5em] 
	{I_6} &= \int_{l} 1 d{\mathbf{r'}}{\rm{ = }}{l_2 } - {l_1},\\[0.5em] 
	{I_7} &= \int_{l} {\left|\vec {l}\ \right|} d{\mathbf{r'}}{\rm{ = }}\frac{{1}}{2}[{({l_2 })^2} - {({l_1})^2}],\\[0.5em] 
	{I_8} &= \int_{l} {\vec l \cdot \vec l} d{\mathbf{r'}}{\rm{ = }}\frac{1}{3}[{({l_2 })^3} - {({l_1})^3}].
\end{align}
Based on the above eight integrals, we can accurately calculate the nearly singular and singular integrals in (\ref{LL}) and (\ref{KK}) in the proposed SS-SIE formulation.

\end{document}